\documentclass[sigconf,screen]{acmart}
\AtBeginDocument{%
  \providecommand\BibTeX{{%
    \normalfont B\kern-0.5em{\scshape i\kern-0.25em b}\kern-0.8em\TeX}}}

\setcopyright{acmlicensed}
\copyrightyear{2024}
\acmYear{2024}
\acmDOI{10.1145/3658644.3690209}

\acmConference[CCS '24]{ACM Conference on Computer and Communications Security}{October 14--18,
  2024}{Salt Lake City, USA}
\acmISBN{978-1-4503-XXXX-X/18/06}

\usepackage{tikz}
\usepackage{amsmath}

\usepackage{blindtext}

\usepackage{caption}
\usepackage{paralist}
\usepackage{subcaption}

\usepackage{lipsum}  

\usepackage{balance}

\usepackage{multicol}
\usepackage{siunitx}
\usepackage{url}
\usepackage{pifont}
\usepackage{cleveref}
\usepackage{xstring}
\usetikzlibrary{shapes, arrows,positioning,chains,backgrounds,fit,matrix}

\usepackage{etoolbox}

\usepackage[frozencache]{minted}
\usepackage{listings}

\usepackage{adjustbox}

\usepackage{fontawesome5}

\usepackage[framemethod=TikZ]{mdframed}

\newmdenv [%
    roundcorner=5pt
    outerlinewidth=0pt,
    innerlinewidth=0pt,
    backgroundcolor=black!10,
    outerlinecolor=black,
    linecolor=black,
    nobreak=false,
    skipabove=0.8em,
    skipbelow=0.7em,
]{takeaway}

\renewcommand{\paragraph}[1]{\medskip\noindent\textbf{#1.}\hspace{1ex minus .1ex}}

\newcommand{\circleone}{\ding{202}\xspace}
\newcommand{\circletwo}{\ding{203}\xspace}
\newcommand{\circlethree}{\ding{204}\xspace}
\newcommand{\circlefour}{\ding{205}\xspace}
\newcommand{\circlefive}{\ding{206}\xspace}
\newcommand{\circlesix}{\ding{207}\xspace}
\newcommand{\circleseven}{\ding{208}\xspace}

\newcommand{\new}[1]{#1}

\newcommand{\eg}{e.g.,\xspace}
\newcommand{\ie}{i.e.,\xspace}

\newcommand{\titletoolname}{DarthShader\xspace} %
\newcommand{\toolname}{\textsc{Darth\-Shader}\xspace}
\newcommand{\toolnameNoSeeds}{\textsc{Darth\-Shader-{\rlap{\ }}-}\xspace}
\newcommand{\naga}{\textsc{naga}\xspace}
\newcommand{\wgslc}{\textsc{wgslc}\xspace}
\newcommand{\tint}{\textsc{tint}\xspace}
\newcommand{\dxc}{\textsc{dxc}\xspace}
\newcommand{\wgslsmith}{\textsc{wgslsmith}\xspace}
\newcommand{\astfuzzer}{\textsc{astfuzzer}\xspace}
\newcommand{\libafl}{\textsc{libAFL}\xspace}
\newcommand{\regexfuzzer}{\textsc{regexfuzzer}\xspace}
\newcommand{\wgslgenerator}{\textsc{wgslgenerator}\xspace}

\newcommand{\artifacturl}{\url{https://github.com/wgslfuzz/darthshader}}
\newcommand{\zenodourl}{\url{https://doi.org/10.5281/zenodo.13302737}}

\newcommand{\numbugs}{\new{39}\xspace}
\newcommand{\numcves}{\new{15}\xspace}

\newcommand{\firefox}{Firefox\xspace}
\newcommand{\chrome}{Chrome\xspace} 
 
\newcommand{\firefoxlogo}{\faIcon{firefox-browser}\xspace}
\newcommand{\chromelogo}{\faIcon{chrome}\xspace} 
\newcommand{\safarilogo}{\faIcon{safari}\xspace}

\newcommand{\webgpu}{WebGPU\xspace}
\newcommand{\wgsl}{WGSL\xspace}

\newcommand{\bugreportchromium}[1]{\href{https://issues.chromium.org/issues/#1}{\color{black}{chromium~#1}}}
\newcommand{\bugreportcve}[1]{\href{https://www.cve.org/CVERecord?id=#1}{\color{black}{#1}}}
\newcommand{\bugreporttint}[1]{\href{https://bugs.chromium.org/p/tint/issues/detail?id=#1}{\color{black}{tint~#1}}}
\newcommand{\bugreportnaga}[1]{\href{https://github.com/gfx-rs/naga/issues/#1}{\color{black}{naga~#1}}}
\newcommand{\bugreportwgpu}[1]{\href{https://github.com/gfx-rs/wgpu/issues/#1}{\color{black}{wgpu~#1}}}
\newcommand{\bugreportwebkit}[1]{\href{https://bugs.webkit.org/show_bug.cgi?id=#1}{\color{black}{webkit~#1}}}

\definecolor{lightgray}{rgb}{.9,.9,.9}
\definecolor{darkgray}{rgb}{.4,.4,.4}
\definecolor{purple}{rgb}{0.65, 0.12, 0.82}

\Crefname{lstlisting}{Listing}{Listings}

\lstdefinelanguage{JavaScript}{
    keywords={typeof, new, true, false, catch, function, return, null, catch, switch, var, if, in, while, do, else, case, break, let, for, async},
    keywordstyle=\color{blue}\bfseries,
    ndkeywords={class, export, boolean, throw, implements, import, this},
    ndkeywordstyle=\color{darkgray}\bfseries,
    identifierstyle=\color{black},
    sensitive=false,
    comment=[l]{//},
    morecomment=[s]{/*}{*/},
    commentstyle=\color{purple}\ttfamily,
    stringstyle=\color{red}\ttfamily,
    morestring=[b]',
    morestring=[b]"
}



\setcopyright{none} %
\renewcommand\footnotetextcopyrightpermission[1]{} %
\begin{document}

\title{\titletoolname: Fuzzing WebGPU Shader Translators \& Compilers}

\author{Lukas Bernhard}
\affiliation{
  \institution{CISPA Helmholtz Center for Information Security}
  \country{Germany}
}
\email{lukas.bernhard@cispa.de}

\author{Nico Schiller}
\affiliation{
  \institution{CISPA Helmholtz Center for Information Security}
  \country{Germany}
}
\email{nico.schiller@cispa.de}

\author{Moritz Schloegel}
\affiliation{
  \institution{CISPA Helmholtz Center for Information Security}
  \country{Germany}
}
\email{moritz.schloegel@cispa.de}

\author{Nils Bars}
\affiliation{
  \institution{CISPA Helmholtz Center for Information Security}
  \country{Germany}
}
\email{nils.bars@cispa.de}

\author{Thorsten Holz}
\affiliation{
  \institution{CISPA Helmholtz Center for Information Security}
  \country{Germany}
}
\email{holz@cispa.de}

\begin{abstract}
A recent trend towards running more demanding web applications, such as video games or client-side LLMs, in the browser has led to the adoption of the WebGPU standard that provides a cross-platform API exposing the GPU to websites. This opens up a new attack surface: Untrusted web content is passed through to the GPU stack, which traditionally has been optimized for performance instead of security. 
Worsening the problem, most of WebGPU cannot be run in the tightly sandboxed process that manages other web content, which eases the attacker's path to compromising the client machine. 
Contrasting its importance, WebGPU shader processing has received surprisingly little attention from the automated testing community. Part of the reason is that shader translators expect highly structured and statically typed input, which renders typical fuzzing mutations ineffective. Complicating testing further, shader translation consists of a complex multi-step compilation pipeline, each stage presenting unique requirements and challenges. 

In this paper, we propose \toolname, \new{the first language fuzzer} that combines mutators based on an intermediate representation with those using a more traditional abstract syntax tree.
The key idea is that the individual stages of the shader compilation pipeline are susceptible to different classes of faults, requiring entirely different mutation strategies for thorough testing. 
\new{By fuzzing the full pipeline, we ensure that we maintain a realistic attacker model.}
In an empirical evaluation, we show that our method outperforms the state-of-the-art fuzzers regarding code coverage. Furthermore, an extensive ablation study validates our key design.
\toolname found a total of \numbugs software faults in all modern browsers---Chrome, Firefox, and Safari---that prior work missed. For \numcves of them, the Chrome team assigned a CVE %
, acknowledging the impact of our results.

\medskip
\noindent{}\emph{This is the author's version of the work. It is posted here for your personal use. Not for redistribution. The definitive Version of Record was published in ACM Conference on Computer and Communications Security (CCS), \url{http://dx.doi.org/10.1145/3658644.3690209}.}

\end{abstract}



{
\maketitle
}

\section{Introduction}%
\label{sec:Introduction}

The internet and the web have been game changers in the past decades, enabling instant access to global news, constant connection with friends and acquaintances, and many types of new business models.
Web browsers, in particular, play a crucial role in this ecosystem, as they are the most important applications to access the web for many users.
However, the ubiquitous connectivity of the internet also enables adversaries with malicious intent, exposing users to potential threats as they navigate the web. 
A common security risk is memory safety violations~\cite{thomas2019proactive}, which have been the starting point for many successful attacks in the past.

As a result, we require fundamental, proactive measures to improve defenses against such threats and strengthen web browsers against various attack vectors. 
By using hardware-supported security features such as memory randomization (ASLR) and non-executable memory regions, web browsers can reduce the risk of exploits that attempt to execute arbitrary code.
Moreover, rigorous testing needs to be performed on all browser components. This includes web APIs~\cite{dominowebidlfuzzer,dinh2021favocado} and JavaScript engines~\cite{gross2023fuzzilli,wang2019superion,park2020fuzzing,salls2021token,han2019codealchemist}, given that they are often targeted due to their complexity and the fine-grained control they expose to adversaries.
In addition, \emph{sandboxing} is a crucial defense mechanism designed to prevent code from performing malicious actions or accessing sensitive data outside its intended scope~\cite{urlChromeSandbox,urlFirefoxSandbox}.
This technique enforces a strict separation between the content of different websites in different processes (called \emph{site isolation}~\cite{reis2009isolating}) and most importantly between web content and the privileged components of the browser, \eg those with access to the file system. 
Technically speaking, sandboxing is implemented by executing code of different sites in separate processes with restricted authorizations. 
Each process is confined by a security policy enforced at the operating system level, which specifies the system resources and cross-process communication channels it has access to.

At the same time, the practical requirement and drive for better performance in web applications, particularly in graphics-intensive areas such as online gaming and video streaming, created a significant demand for improved graphics processing capabilities in browsers. One important development in this area is the recent introduction of \emph{WebGPU}~\cite{urlGPUwebW3C}, a cross-platform API that enables web content to access the computational resources of GPUs. 
Part of this API are WebGPU shaders, essentially small programs that run on the GPU to perform complex rendering operations and general-purpose computations efficiently. 
With WebGPU, these shaders are provided by a website and then processed by a dedicated software component in the browser.
This component is responsible for compiling the shader code into a lower-level, operating system-specific format intended for execution by the GPU.
For instance, in a Windows environment, the shaders would be translated into DirectX bytecode.

Unfortunately, exposing GPU interfaces to web content leads to new attack vectors. For example, in Mozilla Firefox and Google Chrome, the GPU process uses a less strict sandbox for shader processing on Windows, rendering it an attractive target. On some operating systems, the GPU process works \emph{without} a sandbox, further increasing the risk of handling untrusted shader programs.
Consequently, input controlled by attackers ends up in a browser process not protected by a strong sandbox, posing a potential security
risk.
Given the critical nature of shader compilers in the graphics pipeline and their exposure to external, untrusted inputs, one would expect rigorous testing to ensure their security and reliability. 
Contrary to this expectation, we found a significant gap in shader compiler testing in both the literature and industry practices. 
While other browser components have been extensively researched and tested, the testing of shader compilers has been largely insufficient and ineffective, posing a security risk that undermines the primary defense mechanism of sandboxing.
Prior attempts, such as \regexfuzzer~\cite{donaldson2023industrial} and \astfuzzer~\cite{donaldson2023industrial} require a high-quality seed corpus for shader fuzzing and lack intermediate representation (IR) level mutations, limiting their effectiveness in modifying complex data structures and control flows. 
On the other hand, generative testing methods such as \wgslsmith~\cite{mohsin2022wgslsmith} and \wgslgenerator~\cite{wgslgenerator}, which are comparable to Csmith~\cite{yang2011finding} and Xsmith~\cite{Hatch23Generating}, do not implement seed-based mutations, resulting in insufficient branch coverage and high rejection rates, as our empirical evaluation in ~\Cref{sec:eval-semantic} shows.

In this paper, we address this problem and present the design and implementation of \toolname, a fuzzing framework specifically tailored to effectively test the WebGPU stack in modern browsers for memory safety violations and shader compiler errors.
We implemented a generator that fulfills two primary functions: 
On the one hand, it generates a semantically correct input corpus that adheres to the language specification of shaders to enable a deeper exploration of the target under test beyond basic error handling. 
On the other hand, it improves the fuzzing process by injecting new code into an available seed corpus, thus expanding the mutation space by a more extensive variety of expressions and instructions. 
\new{These capabilities make \toolname the first fuzzer with a fully statically-typed IR \emph{and} the capability of both generating and mutating input. This contrasts the existing state of the art: Fuzzilli~\cite{gross2023fuzzilli} has limited static type information (due to the dynamically-typed nature of JavaScript), and other IR-based fuzzers~\cite{chen2021one} can either only generate inputs or mutate them. Generation-based methods~\cite{wgslgenerator,mohsin2022wgslsmith,yang2011finding} cannot leverage coverage feedback, while mutation-based ones~\cite{chen2021one} cannot meaningfully expand the current sample (hence they cannot add entirely new expressions). In contrast, \toolname can correctly infer types for shaders, expand the sample by adding new expressions, and rely on coverage guidance for target exploration.}
In a second step, the shaders are converted into Abstract Syntax Tree (AST) and Intermediate Representation (IR) formats to enable domain-specific mutations.
We designed two sets of mutations since they complement each other: While AST-based mutations test the robustness of the browser's parser and lexer, which are the primary components interfacing with untrusted shader inputs, IR-based mutations target the translation and compilation phases of the shader processing pipeline.
\new{Prior work in language fuzzing has either used mutations on the AST or the IR level. \toolname is the first to combine both in a single tool, relying on their unique advantages to effectively test the shader pipeline.}
The resulting shaders are then sent to the WebGPU shader pipeline, where they test the complete processing stack, \ie both the front-end of the shader processing and the back-end components provided by the actual OS-specific graphics library.
\new{Testing the complete stack ensures that we maintain a realistic attacker model: The back-end in particular makes specific assumptions on the input format. Fuzzing the back-end on its own may uncover various bugs; however, most of them cannot be triggered from the web, which makes them uninteresting for adversaries and vendors alike. Fuzzing the \emph{full} processing pipeline ensures that all input reaching the back-end can be controlled by an adversary and that any bugs found are therefore security-relevant.}

We implemented a prototype of \toolname and tested the state-of-the-art web browsers Chrome, Firefox, and Safari. Our experiments show that our approach succeeds in uncovering on average 11\% and up to 24\% more branch coverage than existing methods. At the same time, \toolname uncovered a total of \numbugs bugs in all shader translators used in all modern web browsers. Furthermore, we uncovered several critical security flaws in the Windows shader compiler \dxc that can be triggered by remotely served web content.
We responsibly disclosed the found vulnerabilities to the vendors and worked together with them to address the identified bugs.
Acknowledging the severity of our findings, Google has assigned \numcves CVEs so far to our reports and awarded a bug bounty for our efforts.

\paragraph{Contributions} In summary, the three main contributions of our work are as follows:
\begin{itemize}

    \item \new{\textbf{Novel IR properties:} \toolname is the first language fuzzer that features both a fully statically-typed IR and the capability to generate and mutate input. %
    This way, our fuzzer can correctly infer shader types, add new expressions to samples, and drive its exploration via coverage feedback.}
    \item \new{\textbf{Novel combination of mutations:} \toolname combines IR and AST mutations, which have individual advantages. IR mutations allow for correct type inference, while mutating the AST allows the generation of inputs that violate the \wgsl grammar.} %
    \item \new{\textbf{Attacker capabilities:} We are the first to fuzz the entire shader pipeline and maintain a realistic attacker model. Overall, our approach uncovered \numbugs bugs in \emph{web-exposed} browser components in Chrome, Firefox, and Safari.}
\end{itemize}

To foster further research on this topic, we release the source code of \toolname at \artifacturl \newline and evaluation artifacts at \zenodourl.

\section{Background}%
\label{sec:Background}
This section gives a short introduction to technical details to understand the relationship between browser sandboxing, WebGPU, the WebGPU Shading Language, and fuzzing.

\subsection{Browser Sandboxing}

Modern browsers like \firefox and \chrome employ a sophisticated multi-process sandboxing model to protect against potentially harmful web content. This approach splits a browser into various processes, each with specific privileges and access rights. For instance, \chrome has a single trusted broker process and multiple untrusted renderer processes~\cite{urlChromeSandbox}. \firefox implements its respective counterparts with one parent process and multiple content processes~\cite{urlFirefoxSandboxModel}. Throughout this discussion, we use the terminology from Chrome while also referring to the respective counterpart in \firefox.

The broker process is the main browser process, running without sandbox restrictions and full user privileges. 
For logical reasons, the broker process does not directly handle untrusted web content. Instead, web content is processed in renderer processes. In contrast to the broker process, renderer processes are forced to request access to system resources via the main process, which mediates these resource requests.
For such requests, the privileged processes evaluate whether the sandboxing policy allows access to the requested resource.
The primary defense against compromised renderer processes is their limited access to system resources, which are meticulously controlled and mediated by the broker process. %
This split-privilege model is a defense-in-depth mechanism, effectively restricting access from compromised renderer processes to the host system by isolating different browser components. For example, an attacker exploiting a vulnerability in the JavaScript engine still needs an additional exploit to escape the sandbox. %

The most common type of sandboxed processes is the renderer process, which processes the majority of web content. The sandbox of the renderer process is the most restrictive, with the minimal set of privileges required to execute the specific web content.
Notably, the kernel attack surface is reduced by blocking access to Windows' graphics subsystem \textsc{win32k.sys}, a component historically plagued by security vulnerabilities.
Some web content, including WebGPU, has a legitimate need to access the OS graphic subsystem. As the renderer sandbox prevents direct access to graphics resources, such resource requests must be outsourced to a process not confined by tight sandboxing rules.
The process that handles graphics subsystem requests is called the GPU process. Compared to the renderer process, the GPU process has a much larger kernel API surface, including access to \textsc{win32k.sys}. Albeit not running with full user privileges as the main process, the sandboxing rules confining the GPU process are much less restrictive. %
An IPC channel with multiple renderer processes and a less restrictive sandbox makes the GPU processes an interesting target, as escaping the sandbox of the GPU process is easier due to the increase in attack surface.

\begin{figure}
   \centering
   \graphicspath{{figures}}
   \def\svgwidth{\columnwidth}
   \begin{scriptsize}
       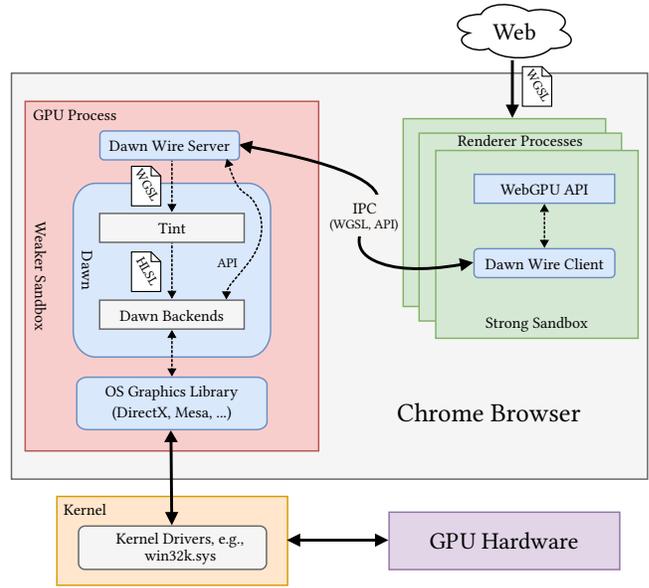
   \end{scriptsize}
    \caption{High-level overview of the multi-process model of Chrome with a focus on the components relevant for WebGPU and shader translation.}
    \label{fig:chrome_gpu_sandboxing}
\end{figure}

\subsection{WebGPU}
\webgpu is a new API designed to expose the functionality of modern  graphic APIs like Vulkan, Metal, and Direct3D to the web. The \webgpu standard effectively supersedes the predecessor WebGL JavaScript API, which primarily mirrors the older OpenGL ES~\cite{urlGPUwebW3C} API. The goal of \webgpu is to allow richer and more complex graphics applications to run portable on the web while providing access to the graphics and computing capabilities of modern GPU hardware. In contrast to WebGL, \webgpu separates the resource management, work preparation, and submission to the GPU~\cite{urlWebGPUfirefox, urlMetalShadingLang}. 

Furthermore, \webgpu offers a low-level interface that allows developers fine-grained control over GPU resources and operations to provide a more efficient access to the underlying hardware. Another design constraint for \webgpu is imposed by the sandboxing architecture, implemented in modern browsers using a GPU process. The browser runs a single process responsible for GPU interaction, communicating with the renderer processes through IPC~\cite{urlGPUwebExplainer}. %

A high-level overview of the different abstraction layers of \webgpu within the browser and the relationship to the operating system is shown in~\Cref{fig:chrome_gpu_sandboxing}.
\webgpu is exposed to web content via a standardized interface. API requests and shader invocations, generally, cannot be fulfilled by the renderer process due to sandboxing policies. Instead, API requests and shaders are passed via IPC to the GPU~Process. In the GPU~Process, API requests are sanitized and forwarded to the Dawn~Backend. This back-end is essentially an abstraction layer of the OS-specific graphics library, such as DirectX on Windows and Mesa on Linux.
As the shader language of \webgpu is unknown to native graphics back-ends, the shaders have to be translated into an appropriate format. For example, on Windows shaders are translated from the platform-independent \webgpu shader language into HLSL (High-Level Shading Language)~\cite{urlHSLSMicrosoft}. Since the shader code is entirely controlled by third parties, the shaders have to be considered untrusted. The Chrome component implementing this translation, called \tint, will be explored in more detail later.
Once the shaders are translated, they are forwarded to the native back-end alongside other API requests. The native back-ends typically consist of a userland component running in the address space of the GPU processes.
Noteworthy, despite the shaders passed to the native back-end being generated by \tint, adversaries still exercise a high degree of control. This is a consequence of the shader translator having to preserve the semantics of the input shader, including input/output behavior, loop structures, and function calls.
Once the user-mode part of the shader compilation finishes, compiled shaders are passed to the kernel, eventually reaching the GPU hardware.

\subsection{\wgsl \& Shader Translation}
The \webgpu Shading Language, also known as \wgsl, is the shading language utilized in \webgpu. The language enables developers to write shaders, which are small programs executed on the GPU that define how graphical elements are rendered in web applications, including tasks like lighting, texturing, and effects. The \wgsl coding style closely resembles that of Rust; a code example can be found in~\Cref{fig:shaderAbstractionLayersText}. \wgsl is statically typed and designed to be similar to other shading languages like MSL (Metal Shading Language)~\cite{urlMetalShadingLang} and HLSL. \wgsl provides features necessary for modern graphics programming while being tailored specifically for the \webgpu API.
The language is closely integrated with the \webgpu API, allowing shaders to interact with other parts of the rendering pipeline, such as vertex stages, and communicate with buffers and textures.

Shader translation is a crucial step in the graphics rendering process. \webgpu shaders, written in the shading language \wgsl, are not directly accepted by any graphics back-end such as DirectX or Metal. Hence \wgsl shaders must undergo translation into a platform-specific shader format, such as SPIR-V, HLSL, or MSL. The translation from \wgsl to an OS-specific format ensures compatibility across different OS-specific graphics back-ends. In \firefox, the \naga~\cite{nagalib} component handles the translation from \wgsl to the OS-specific shader format. Similarly, in \chrome, the \tint~\cite{tintGoogle} component fulfills this role.
The shader translators are the first component processing \wgsl shaders; the renderer process generally treats the shader as a blob. Hence the translators are the first component being exposed to potentially malicious shaders. As shown in~\Cref{fig:chrome_gpu_sandboxing}, this processing occurs outside of the tightly sandboxed renderer process.
Once the shaders have been transformed into an OS-specific format, they are passed down to the graphics back-end provided by the OS. On Windows, this native back-end is DirectX, which consumes shaders in the HLSL format. DirectX first translates HLSL into LLVM IR and subsequently runs various optimization passes. This entire optimization process runs in the GPU process of the browser, in a component called \dxc. Once optimization is complete, \dxc emits bitcode intended for the GPU-specific kernel driver.

\subsection{Language Fuzzing}
Fuzzing is a software testing technique used to discover bugs in software systems. It involves providing invalid, unexpected, or random data as inputs to a program and observing its behavior. These inputs are produced using either mutation operations or generational methods.
In cases where the software processes binary file formats, typical mutations might include bit-flipping or inserting specific integer values~\cite{aschermann2019redqueen,wang2010taintscope}.
In contrast to targets consuming a binary file format, language processors pose a different set of challenges. Here, inputs are expected to adhere to the rigid rules of a grammar or language specification. Thus, traditional mutations are often ineffective for language processing because they prevent the program from parsing the input correctly. Such inputs lead to an early exit of the target. %
To address this issue, fuzzers for language inputs usually work with an abstract syntax tree (AST)~\cite{holler2012fuzzing, aschermann2019nautilus,wang2019superion,han2019codealchemist,wang2017skyfire,veggalam2016ifuzzer}. Instead of flipping bits, they apply tree-edit operations to the AST, allowing for more sophisticated manipulation that respects the structure of the language, leading to more effective testing.
Successfully applying these AST mutations to statically typed languages is not trivial. Consider applying mutation operations to a dynamically typed language such as JavaScript. Replacing the inputs of an addition operation will exercise significant portions of a JavaScript engine, even if the mutation results in a runtime error during script execution.
In contrast, in statically typed languages, mutations that break the static typing rules result in early exits of the tested application.
This premature termination prevents the fuzzer from exploring deeper and potentially vulnerable code paths.

\section{Design}%
\label{sec:Design}
\begin{figure}[tb]
   \centering
   \graphicspath{{figures}}
   \def\svgwidth{\columnwidth}
   \begin{small}
       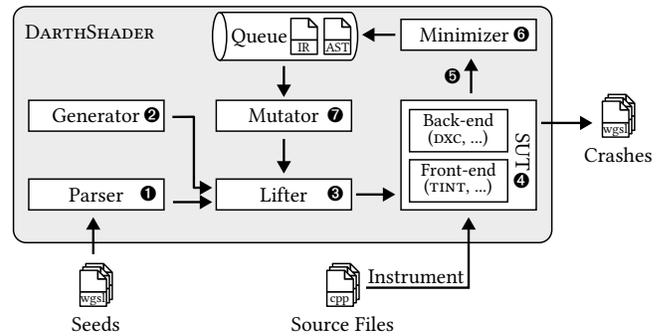
   \end{small}
   \caption{High-level overview of \toolname, showing the relationship between fuzzer components and the SUT.}%
   \label{fig:overview}
\end{figure}

To effectively test shader translators and shader compilers, we introduce a new approach, \toolname. We commence with a description of the overall design and how its components interact. Then, we discuss how inputs are represented on the AST and the IR layer. Finally, we explore key components in more detail, explaining how they operate on the two input representations.

A high-level overview of our approach is shown in~\Cref{fig:overview}. As an optional first step \circleone, we import existing shader files as seeds. While this is common for binary file format fuzzers, some language fuzzers do not support this capability~\cite{aschermann2019nautilus}. 
During parsing, \toolname translates all provided shaders into an AST representation as well as an IR representation. 
In addition to seeds, the initial corpus comprises samples emitted by our seed generator \circletwo. 
Analog to imported seeds, generated samples are stored in an AST and an IR representation, not in a textual representation.
The following section outlines our motivation for including two distinct representations.
Because these AST and IR representations are internal to \toolname, they must be converted~\circlethree to a textual representation before passing them to the system-under-test (SUT). When converting an AST, simply unparsing the tree yields a \wgsl shader, which is the input expected by shader translators. Transforming the IR to text is a more intricate process; on a conceptual level, we lift the IR to \wgsl, resulting in a \wgsl shader as well.
Once lifting completes, we pass the shader to the SUT~\circlefour. 
Interestingly, our SUT can consist of up to two components: Our primary and immediate targets are shader translators, which modern web browsers use to process \wgsl shaders.
Additionally, the SUT may include a back-end compiler, such as \dxc. If this is the case, our fuzzing input is first processed by the shader translator, functioning as the front-end, before the translator's output is then in turn passed to the back-end shader compiler. 
This approach allows us to test both the shader translator and compiler simultaneously.
It is noteworthy that not all our inputs necessarily reach the back-end, as the front-end may discard inputs, for example, when they cannot be parsed.
As is typical for fuzzing, we prepare the SUT by compiling it with coverage feedback instrumentation and ASAN for sanitization before commencing the fuzzing phase.
Executing the instrumented target application with a shader can lead to one of three outcomes: If the application crashes, we save the relevant sample for manual inspection. If an execution reaches new code paths, we keep those samples for further processing~\circlefive. All other samples are discarded.
The samples we retained for achieving novel coverage are sent to a minimizer~\circlesix, which iteratively reduces the samples by removing parts that do not cover new edges. Once minimization is complete, the reduced sample is added to the queue, which contains all samples that contribute to additional coverage.
From this queue, the mutator~\circleseven selects an input and transforms it based on the available mutations. The set of available mutations depends on the type of selected sample. More precisely, IR samples undergo a set of IR mutations whereas AST samples are modified via tree-edit operations.

\begin{figure*}[tb]
\begin{subfigure}[b]{0.32\linewidth}
\begin{minted}[%
    frame=lines,
    framesep=2mm,
    fontsize=\footnotesize,
    stripnl=false,
    xleftmargin=0.8em,
    xrightmargin=0.0em,
]{rust}
struct VertexOut {
  @builtin(position) pos: vec4<f32>,
  @location(0) col: vec4<f32>
}

fn color() -> vec4<f32> {
  return vec4<f32>(0, 0, 0, 0);
}

fn vert_main(@location(0) pos: vec4<f32>)
        -> VertexOut {
  var out: VertexOut;
  out.pos = pos;
  out.col = color();
  return out;
}
\end{minted}
\caption{Textual representation of \wgsl shader}%
\label{fig:shaderAbstractionLayersText}
\end{subfigure}
\hfill
\begin{subfigure}[b]{0.32\linewidth}
  %  \definecolor{codeblue}{HTML}{0000FF}
  %  \definecolor{coderedish}{HTML}{B0003f}
  %  \definecolor{codegreen}{HTML}{007F00}
  %  \definecolor{codegrey}{HTML}{666666}
  %  \begin{tiny}
  %  \begin{tikzpicture}[>=stealth, every node/.style={rectangle, draw, minimum size=0.2cm, rounded corners=.05cm}]
  %   \graph [tree layout, grow=down, fresh nodes, level distance=0.23in, sibling distance=0.25in]
  %       {
  %           decl -> {
  %             function -> { "\textbf{\textcolor{codegreen}{\texttt{fn}}}", id -> { "\textcolor{codeblue}{\texttt{color}}" }, "\texttt{(}", "\texttt{)}", template -> { id -> { "\textbf{\textcolor{codeblue}{\texttt{vec4}}}" }, "list" -> { id -> { "\textbf{\textcolor{coderedish}{\texttt{f32}}}" } } } },
  %             compound -> { "\texttt{$\lbrace$}", stmt -> { return -> { "\textbf{\textcolor{codegreen}{\texttt{return}}}", call -> { list -> { id -> "\textcolor{codegrey}{\texttt{0}}", id -> "\textcolor{codegrey}{\texttt{0}}", id -> "\textcolor{codegrey}{\texttt{0}}", id -> "\textcolor{codegrey}{\texttt{0}}" } } }, "\texttt{;}" } , "..." }
  %           }
  %       };
  %   \end{tikzpicture}
  %   \end{tiny}
    \includegraphics{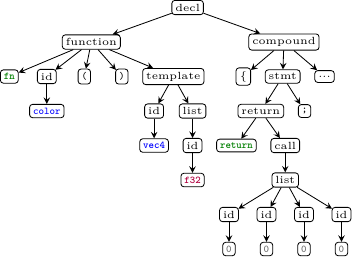}
    \vspace{0em}%
   \caption{AST representation}%
   \label{fig:shaderAbstractionLayersAST}
\end{subfigure}
\hfill
\begin{subfigure}[b]{0.33\linewidth}
\begin{minted}[%
    frame=lines,
    framesep=2mm,
    fontsize=\footnotesize,
    stripnl=false,
    xleftmargin=0.8em,
    xrightmargin=1.8em,
]{hlsl}
Expressions:                // Types:
[1]: FunctionArgument(0)    // vec4<f32>
[2]: LocalVariable([1])     // VertexOut
[3]: Access { [2], idx: 0 } // vec4<f32>
[4]: Access { [2], idx: 1 } // vec4<f32>
[5]: CallResult([1])        // vec4<f32>
[6]: Load { ptr: [2] }}     // VertexOut

Statements:
EmitExpr([3])
Store { pointer: [3], value: [1] }
EmitExpr([4])
Call { fun: color, args: [], res: [5] }
Store { pointer: [4], value: [5] }
EmitExpr([6])
Return { value: Some([6]) }
\end{minted}
\caption{IR representation}%
\label{fig:shaderAbstractionLayersIR}
\end{subfigure}
\caption{Multiple representations of a shader. (a) Source code, as processed by the browser. This format is amenable to byte-level mutations only. (b) An excerpt from the shader, parsed into an AST. This format supports tree-based mutations, such as swapping nodes. (c) The IR. This format facilitates domain-specific mutations, such as altering function prototypes.}
\label{fig:shaderAbstractionLayers}
\end{figure*}

\subsection{Language Representation}
When designing a fuzzer, one key decision is selecting the abstraction layer at which mutations will be applied. This choice can be straightforward for binary file formats, but it is more complex for language fuzzers due to a larger design space.
In~\Cref{fig:shaderAbstractionLayers}, we depict three abstraction layers commonly encountered in language fuzzers. On the left side, \Cref{fig:shaderAbstractionLayersText} illustrates the source code of a shader program. This is the input representation as processed by the SUT. However, text is not well-suited for mutations commonly used in language fuzzing. 
Instead, language fuzzers typically represent inputs as either an AST or IR, shown in~\Cref{fig:shaderAbstractionLayersAST} and \Cref{fig:shaderAbstractionLayersIR}. In the following, we discuss the respective advantages and downsides of these two abstraction layers.

\paragraph{Abstract Syntax Tree (AST)}
One common choice~\cite{aschermann2019nautilus,wang2019superion,han2019codealchemist,wang2017skyfire,veggalam2016ifuzzer} for representing inputs in language fuzzing are ASTs, since trees are straightforward to mutate via tree-edit operations. Furthermore, AST mutations allow rigorous testing of lexers and parsers in the SUT, e.g., adding reserved keywords or inserting literals that exceed standard sizes (such as a number that cannot be represented by 64 bits).
However, implementing mutations on ASTs also has downsides. Consider a mutation that changes function \texttt{color()} from~\Cref{fig:shaderAbstractionLayersText} to \texttt{color(a: vec4<f32>)}, i.e., adds a function argument of type \texttt{vec4<f32>}. Implementing such a mutation on the AST representation shown in~\Cref{fig:shaderAbstractionLayersAST} is cumbersome.
Not only do we have to add the argument, but we also need to find all callers of the mutated function and extend the parameter lists with a variable of the correct type. While not an impossible task, the AST representation is unsuitable for such mutations.

\paragraph{Intermediate Representation (IR)}%
\label{sec:design-ir}
In contrast, the IR representation shown in~\Cref{fig:shaderAbstractionLayersIR} allows certain mutations, such as the aforementioned extension of a function call with an additional argument with ease. Not only do we precisely know all callers of \texttt{color()}, we also know the type of all expressions in scope at the respective call site.
Moreover, other complex mutations, such as inserting new types or changing the scope of an expression, are straightforward to implement.
These advantages of IR representations contributed to their usage~\cite{chen2021one,gross2023fuzzilli} in fuzzing. Unfortunately, this abstraction layer is not a panacea for all challenges in language fuzzing. The advantages of AST mutations are aspects for which the IR representation falls short. For example, an IR internally representing numbers as 64~bit cannot emit an oversized literal, as such a number is simply unrepresentable in the IR.

\paragraph{Dual Approach}
Rather than choosing between an AST and an IR representation, \new{one insight of our design} is to incorporate both levels of abstraction. \new{This makes our fuzzer the first that can} harness the specific benefits of each, without the limitations typically associated with either.
As part of our evaluation in~\Cref{sec:eval-ablation}, we show that mutating inputs on an AST layer is well-suited for exploring front-end translators such as \tint. At the same time, an IR approach excels in testing back-end translators such as \dxc. In order to thoroughly stress-test the entire shader translation pipeline, we posit that including both representations is essential and evaluate this insight in an ablation study (see Section~\ref{sec:eval-ablation}).

\subsection{Key Components}
With the overview shown in~\Cref{fig:overview} in mind, we present key components in greater detail. \new{A key feature distinguishing our design from existing language fuzzers is that it uses a statically-typed IR and has the capability to both \emph{generate} and \emph{mutate} inputs.} In the following, we first describe a mechanism for generating \wgsl samples~\circletwo. Next, we explain the methods for mutating these samples at both the IR and AST layers~\circleseven, and we conclude by exploring the minimizer~\circlesix.

\paragraph{Generator \circletwo}%
\label{sec:design-generator}
In absence of informed seeds, the generator produces an initial corpus of inputs. This initial generation aims for a high semantic correctness rate, \ie the majority of samples should conform to the language specification. While a fuzzer should \emph{also} test inputs deviating from the input specification, these violations will be introduced by mutation operations anyway. In contrast, mutation operations typically fail to convert semantically incorrect inputs into ones that meet the required specification. Hence, generating mostly correct inputs as a first step allows for reaching deeper into the target instead of only exploring shallow error handlers. %

To generate a \wgsl program, we first create a pool of types that will be available to this program. Initially, this includes basic scalar types such as \texttt{int32}. We then expand this pool by randomly adding more complex types like structs and vectors. For structs, we randomly choose the number of member variables and select the respective from the pool of types created so far. Once all types are defined, we proceed to create function prototypes and their bodies.
This involves iteratively constructing a list of statements, expressions, and their respective inputs. For example, an if/else statement needs a Boolean condition, which we select from previously generated expressions. If the necessary expression is unavailable, we discard the current statement or expression and try again. This process continues until we reach a predefined limit on function length, ensuring the program generation algorithm terminates.

\paragraph{Mutations \circleseven}%
\label{sec:design-mutator}
One key characteristic of a fuzzer is the set of mutations available for transforming an input to another one. Our design consists of two classes of mutations, IR mutations and AST mutations. 
The first class, \new{IR mutations, are intended to stress-test the later translation and compilation stages that operate on intermediate representations:}

\begin{itemize}
    \item \textbf{Operators:} mutates unary and binary operators, \eg replaces a plus operation with a multiplication.
    \item \textbf{InputReplace:} exchanges the inputs used by expressions and statements.
    \item \textbf{Literals:} replaces literals such as integers and floats either with a random choice or selects from a list of interesting integers, such as powers of 2. 
    \item \textbf{Built-ins:} exchanges calls to built-in functions with a different built-in function call.
    \item \textbf{Types:} mutates types, \eg by resizing arrays or changing the types of scalar variables.
    \item \textbf{CodeGen:} emits additional code at a random location in the input program.
\end{itemize}

Despite IR mutations testing for a sizable set of potential errors, by design they cannot find some classes of errors in the domain of shader translators. When lifting~\circlethree IR code to WGSL, the resulting AST adheres strictly to the grammar. Hence, an entire subset of bugs~\cite{salls2021token} remains unreachable via IR mutations. In order to stress-test the lexer and parser, we have six AST mutations at our disposal:
\begin{itemize}
    \item \textbf{RecursiveReplace:} recursively inserts a subtree~\cite{aschermann2019nautilus} into itself while accounting for the respective node types. This mutation generates pathological trees with a particularly deep nesting of child nodes.
    \item \textbf{Delete:} removes an AST node and all its children.
    \item \textbf{Replace:} replaces the text of AST nodes with a value from a dictionary containing domain-specific tokens.
    \item \textbf{Splice:} crossover mutation that splices a random AST from the corpus into a second AST.
    \item \textbf{Swap:} reorders the children of a single AST node.
    \item \textbf{Identifier:} replaces an identifier (\eg variable name) with a different identifier used elsewhere in the input.
\end{itemize}

\paragraph{Minimizer \circlesix}%
\label{sec:design-minimizer}
Interesting samples, \ie \wgsl inputs that trigger uncovered edges, are scheduled for minimization before adding them to the corpus. 
This step is essential to prevent unbounded input growth, a common issue resulting from splicing operations and code generation, which increase input size.
Keeping input size small has two significant advantages.
First, smaller inputs require less processing time in the SUT and thus increase throughput.
Second, our goal is to retain samples that uncover new, unexplored edges of the SUT. Keeping inputs small is beneficial because it ensures that future splicing mutations---which combine elements from different inputs---retain these valuable features without being diluted by irrelevant data.

The minimization process consists of two steps. Initially, we identify the edges that are both new and consistently triggered by the input. To this end, we repeatedly execute the input and record the consistently reached edges.
Non-deterministically exercised edges may, e.g., be an artifact of randomized data structures such as hash tables or memory allocators. The requirement for removing non-deterministic edges is imposed by the second step: We successively remove small parts of the input and verify whether we still reach all novel edges. Assuming the set of novel edges contains non-deterministic edges, minimization becomes challenging. Most likely, at least some of the flaky edges are no longer exercised by minimized variants; hence, we fail to remove any input parts.
The specific techniques for minimizing an input depend on its type: For AST inputs, we minimize the tree by pruning nodes. For IR inputs, several strategies are available. Examples include the removal of global variables, expression simplification, and statement deletion.

\section{Implementation}%
\label{sec:Implementation}
We implement our proposed design in a tool called \toolname, amounting to $10,000$ lines of code.  While not building on a domain-specific fuzzer, we do reuse components of \libafl~\cite{fioraldi2022libafl} for general fuzzer housekeeping, \textsc{tree-sitter}~\cite{treesitter} for parsing shaders, and \naga~\cite{nagalib} for its IR.
Below, we highlight two of our building blocks.

\paragraph{\libafl} \toolname uses \libafl components for general fuzzing housekeeping, such as coverage evaluation and communication with the SUT. %
Our implementation uses MOPT~\cite{lyu2019mopt} for scheduling and a power-schedule~\cite{bohme2016coverage} for seed selection. When exploring their respective configuration parameters, we observed a 5\% difference in coverage over 24 hours between the best and worst configuration. We selected the best-performing set of parameters for all following measurements.

\paragraph{\naga} Our implementation leverages the IR exposed by \naga, a shader translator part of \firefox. The IR is designed to express semantics common to graphics shaders in general, not only \wgsl. Furthermore, \naga includes the ability to lift the IR to \wgsl source code, exactly the format consumed by the SUTs.
One complimentary upside of using \naga is its ability to utilize seed files written in SPIR-V and GLSL, two widely used shader languages. This capability enlarges the set of available seed files, increasing bug-finding by importing regression tests of other shader processors.

\section{Evaluation}%
\label{sec:Eval}
To evaluate our approach, we compare \toolname against state-of-the-art fuzzers and domain-specific test-case generators. The two main metrics for comparing the different approaches are code coverage and the semantic correctness rate of produced inputs. Furthermore, we perform an ablation study scrutinizing individual design decisions of \toolname. Finally, we test whether our prototype can uncover previously unknown bugs in components exposed to the web, rendering their security a delicate matter. %

\subsection{Setup}
We first describe the experimental setup used during our evaluation, including the hardware environment, tested fuzzers, and evaluation targets.
For all experiments and ablation studies, we perform 10 repetitions over either 24h or 48h. Each fuzzer and its respective target is pinned to a single CPU core, following general guidelines for fuzzing evaluations~\cite{klees2018evaluating}. 

\paragraph{Hardware Environment}
All experiments were performed on an AMD~EPYC~9654 processor with 755~GB of RAM and a SSD as backing storage.

\paragraph{Target Applications} We evaluate code coverage and semantic correctness rate on four web-exposed targets. Furthermore, our bug finding efforts includes an additional target, \textsc{angle}, not supported by competing fuzzers.
\begin{itemize}
\item \textbf{\tint} (\new{commit} \texttt{3de0f00}), the shader compiler of Chrome supporting compilation from \wgsl to HLSL, SPIR-V, and Metal, the respective shader languages of Windows, Linux, and macOS. Our fuzzing harness for \tint translates a single \wgsl shader to each of the three target languages.
\item \textbf{\dxc} (\new{commit} \texttt{0781ded}), the DirectX shader compiler taking HLSL as input and producing an output format based on LLVM IR. Our fuzzing harness first translates \wgsl shaders to HLSL via \tint and subsequently passes the HLSL code to \dxc, so that setup replicates browser usage.
\item \textbf{\naga} (\new{commit} \texttt{61d779d}), the shader compiler of Firefox supporting compilation from \wgsl to HLSL, SPIR-V, and Metal. Analog to the \tint harness, our \naga harness translates a single \wgsl shader to HLSL, SPIR-V, and Metal.
\item \textbf{\wgslc} (\new{commit} \texttt{ad13d16}), the shader compiler of Safari translating \wgsl to Metal. No other output languages are supported.
\end{itemize}

\paragraph{Fuzzers} In the following, we describe the six fuzzers that are evaluated based on code coverage and semantic correctness rate.
\begin{itemize}%
\item \textbf{\toolname}, our approach implements generation on an IR layer and mutations on both IR and AST layer. In this variant of our fuzzer, we include informed seeds.
\item \textbf{\toolnameNoSeeds}, a variant of \toolname that runs without informed seeds. It produces samples with a combination of generation and mutations.
\item \textbf{\wgslsmith}~\cite{mohsin2022wgslsmith} (\new{commit} \texttt{987ddf1}), a domain-specific generator producing \wgsl shaders. This tool is purely generational and does not support coverage feedback. For our evaluation, we wrapped \wgslsmith with \libafl such that we only store samples increasing code coverage.
\item \textbf{\wgslgenerator}~\cite{wgslgenerator} (\new{commit} \texttt{ffbaad4}), a domain-specific \wgsl generator. Analog to \wgslsmith, we added a \libafl wrapper for storing only samples increasing code coverage.
\item \textbf{\regexfuzzer}~\cite{donaldson2023industrial} (\new{commit} \texttt{3de0f00}), a libfuzzer-based fuzzer integrated in \tint. As this fuzzer and its custom mutations are fully integrated, \tint it the only supported target. This fuzzer highly depends on informed seeds and reaches no noteworthy coverage on uninformed seeds.
\item \textbf{\astfuzzer}~\cite{donaldson2023industrial} (\new{commit} \texttt{3de0f00}), a libfuzzer-based fuzzer integrated in \tint. For the same reason as \regexfuzzer, this fuzzer is evaluated on \tint only. Likewise, this fuzzer highly depends on informed seeds.
\end{itemize}

\paragraph{Seeds} While using informed seeds is trivial for binary fuzzers, not all language fuzzers support this capability. For example, the JavaScript fuzzer Fuzzilli~\cite{gross2023fuzzilli} did not include this feature initially.
However, supporting seeds is beneficial because they allow variant analysis of old bugs and utilizing test cases.
We evaluate the performance of \toolname with and without seeds. Our competitors either require informed seeds (\astfuzzer and \regexfuzzer) or cannot use them at all (\wgslsmith and \wgslgenerator). We use the test cases of \tint containing $7,267$ \wgsl files as corpus.

\begin{table}[tb]
    \centering
    \caption{Median semantic correctness rate in percent and standard derivation on all front-end translators. \dxc is a back-end compiler and therefore excluded from this metric.}%
    \label{tab:semRate}
    \begin{adjustbox}{max width=\columnwidth}
    \begin{tabular}{ l S[table-format = 2.2(1), separate-uncertainty] S[table-format = 2.2(1), separate-uncertainty] S[table-format = 2.2(1), separate-uncertainty] }
        \toprule
        \textbf{Fuzzer} \hspace{1em} & \textbf{\tint} & \textbf{\naga} & \textbf{\wgslc} \\
        \midrule
        \toolname & 14.26(0.76) & 18.13(1.48) & 16.88(0.81) \\
        \toolnameNoSeeds & 12.68(0.95) & 14.49(1.26) & 15.47(1.65) \\
        \wgslsmith & 0.77 & 0.83 & 6.08(0.01) \\
        \wgslgenerator & 0.004 & 0.004 & 0.04 \\
        \regexfuzzer & 6.65(0.10) & \multicolumn{1}{c}{--} & \multicolumn{1}{c}{--} \\
        \astfuzzer & 99.25(0.05) & \multicolumn{1}{c}{--} & \multicolumn{1}{c}{--} \\

        \bottomrule
    \end{tabular}
    \end{adjustbox}
    \begin{flushleft}
    \end{flushleft}
\end{table}

\begin{figure*}[tb]
    \centering
    \graphicspath{{figures/}}
    \def\svgwidth{\linewidth}
    \begin{footnotesize}
        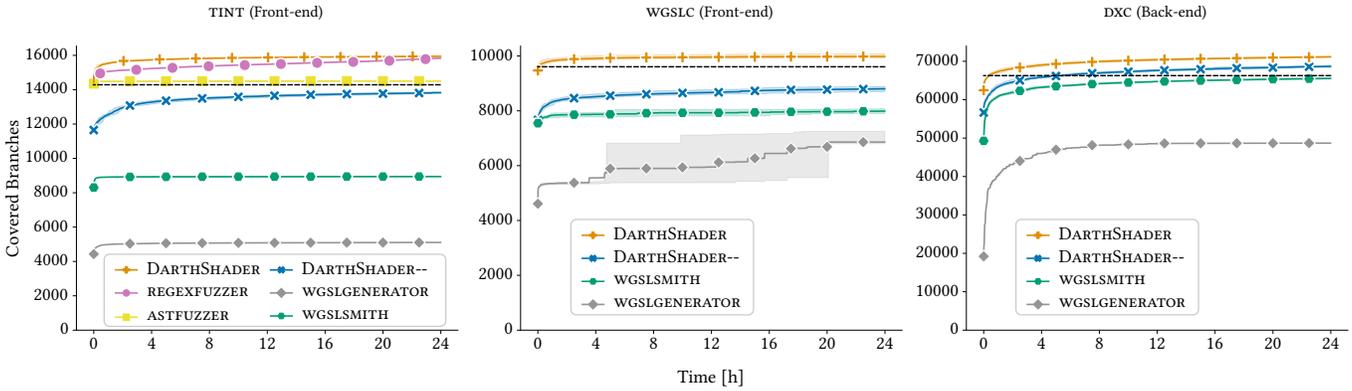
    \end{footnotesize}
    
    \caption{Number of branches covered by running fuzzers over 24h on \dxc, \tint, and \wgslc. Displayed are the median and the 60\% interval of 10 repetitions. The dotted horizontal line shows the coverage inherent in the informed seeds corpus used by \toolname, \astfuzzer and \regexfuzzer. The other fuzzers do not have access to the informed seeds.}
    \label{fig:coverage}
\end{figure*}

\subsection{Semantic Correctness Rate}
The semantic correctness rate quantifies the percentage of samples an SUT processes successfully. Precisely, we assess the proportion of \wgsl shaders a SUT accepts and converts into a back-end-specific output format. Only such translated shaders are forwarded to OS-specific compiler back-ends. Consequently, a correctness rate approaching 0\% is inadequate for testing downstream components, as most inputs are discarded before ever reaching the back-end. On the other hand, a correctness rate of 100\% implies that no invalid inputs are produced, despite providing inputs containing errors can be crucial to detect bugs in the translation step.
\Cref{tab:semRate} shows the correctness rate measured for the front-end translators \tint, \naga, and \wgslc. These rates are based on actual executions rather than queue samples, thus avoiding queue survivor bias.

We observe that \toolname yields a correctness rate of 12\%-18\%. This rate allows for putting significant pressure on front-end translators due to samples violating the specification while also reaching OS-specific back-ends. Noteworthy, a correctness in the range of 12\%-18\% does \emph{not} imply a SUT spends the majority of its processing time in the front-end.
Samples reaching the back-end require more processing time, whereas semantically incorrect samples are rejected quickly. %
\wgslsmith and \wgslgenerator yield a correctness rate $\ll1\%$ across most targets. Notably, both tools exhibit a significantly higher correctness rate with \wgslc, which we suspect results from the less mature state of the validator.
In stark contrast, the correctness rate of \astfuzzer approaches 100\%, indicating that most inputs conform to the specification. However, a correctness rate this high is counterproductive, as completely correct samples exert minimal pressure on the front-end translators.

\begin{takeaway}
A balanced semantic correctness rate is required to test both front-end and back-end translators.
\end{takeaway}

\subsection{Coverage Experiments}%
\label{sec:eval-coverage}
We use coverage as a metric to compare fuzzer performance. For this measurement, we first compile the target applications with llvm-cov~\cite{llvmcov}. Then, we replay the inputs from all fuzzers on the same instrumented binary to derive a fair and consistent comparison.
We use branch coverage on \tint, \wgslc, and \dxc. As an exception, we use line coverage for \naga. The rationale behind this choice is Rust's pervasive usage of pattern matching for diverging control-flow, which does not correspond to a source-code if-else construct. Hence llvm-cov branch coverage does \emph{not} consider pattern matching induced control-flow constructs.
As mentioned earlier, each fuzzer runs 24 hours per target with ten repetitions to account for inherent randomness in the fuzzing process.

\paragraph{Coverage over Time}
\label{sec:eval-semantic}
Branch coverage over time is one of the key performance criteria of fuzzers. We show this metric for three target applications in \Cref{fig:coverage} and line coverage for \naga in \Cref{fig:naga_coverage}. Taking into account only fuzzers without access to informed seeds, \toolnameNoSeeds outperforms its competitors \wgslgenerator and \wgslsmith. Noteworthy, our approach based on generation and mutation achieves almost the same coverage as the vast corpus of informed seeds. Coverage inherent in the seed corpus is marked by the dotted horizontal line.
When comparing \toolname with competitors requiring informed seeds, our approach either yields higher (\astfuzzer) or similar coverage (\regexfuzzer). As the seeds already cover a significant number of branches, our measurements show only a small improvement over the informed corpus.

\begin{figure}[tb]
    \centering
    \graphicspath{{figures/}}
    \def\svgwidth{\linewidth}
    \begin{footnotesize}
        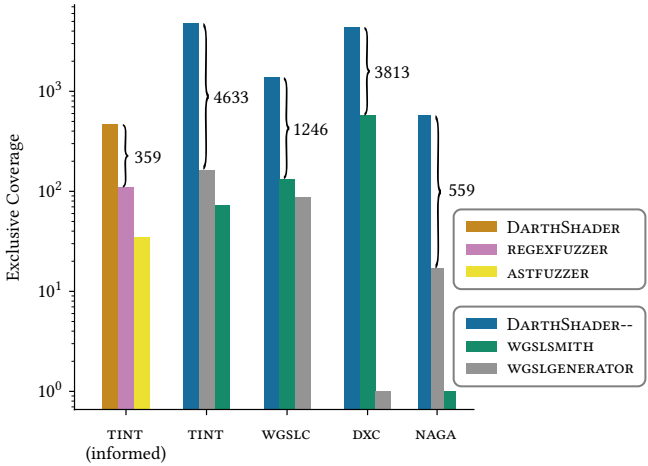
    \end{footnotesize}
    
    \caption{Logarithmic plot showing the branches \emph{exclusively} covered by \toolname and its competitors. For \naga line coverage is used (see ~\Cref{sec:eval-coverage}). To allow a fair comparison, we separate the tools based on access to informed seeds. To derive this metric, we merged the coverage of ten repetitions per fuzzer on each target.}
    \label{fig:unique}
\end{figure}

\paragraph{Exclusively Covered Branches} In addition to coverage over time, we measure the branches exclusively covered by a single fuzzer~\cite{bars2023fuzztruction}, \ie branches that are not covered by competitors. For this measurement, we merge the coverage results of the ten runs per fuzzer and target. For example, we take all ten runs of \toolname on \dxc and compute the number of branches not covered by \wgslsmith and \wgslgenerator.
In order to ensure a fair comparison, this part of the evaluation separates the tools  based on access to seeds. On \tint, we separate the evaluation into two groups, one having access to seeds, whereas the other group does not. On all other SUTs, we compare \toolnameNoSeeds to its competitors because, by design, none of them can utilize seeds.
The results of this evaluation are shown in~\Cref{fig:unique}. On all targets, \toolname and \toolnameNoSeeds cover branches not covered by any other fuzzer.
Particularly interesting are the comparisons between groups of fuzzers without access to an informed seed corpus. The combination of generation and mutation implemented in \toolnameNoSeeds strongly outperforms the competing approaches.

\begin{takeaway}
\toolname outmatches competitors at testing front-end shader translator and back-end compilers.
\end{takeaway}

\begin{figure*}[tb]
    \centering
    \graphicspath{{figures/}}
    \def\svgwidth{\linewidth}
    \begin{footnotesize}
        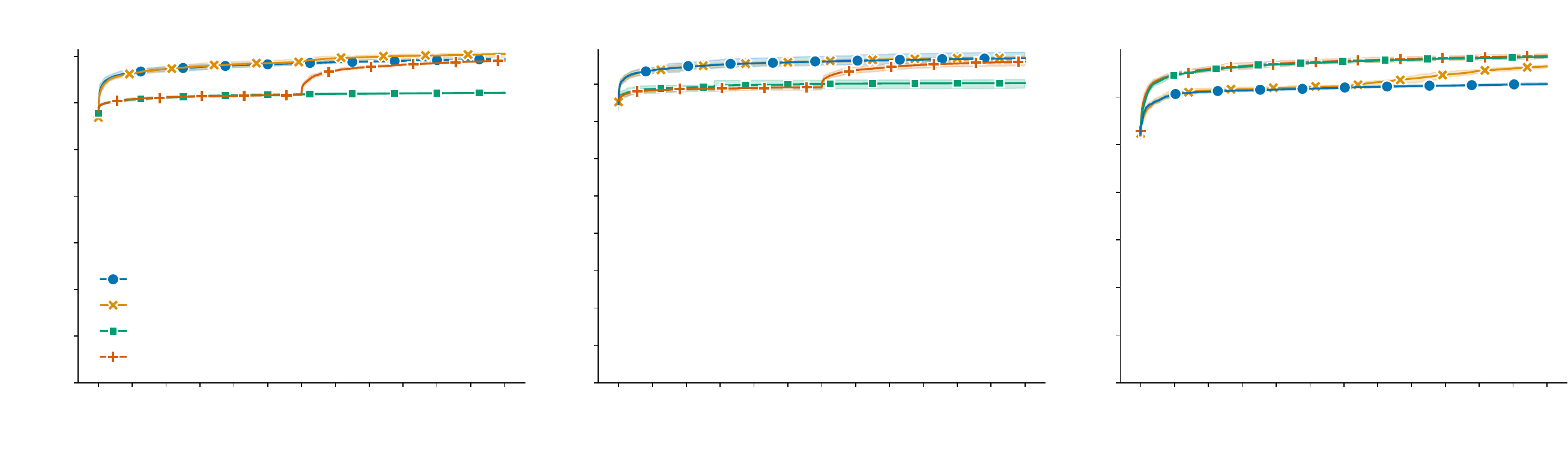
    \end{footnotesize}
    \caption{Ablation study measuring the impact of IR and AST mutations on three shader translators. Most branches are reachable via AST mutations on the front-end translators \tint and \wgslc. In contrast, \dxc is explored much better by IR mutations, with AST mutations contributing little additional coverage.}
    \label{fig:ablation}
\end{figure*}

\subsection{Ablation Study}%
\label{sec:eval-ablation}
To better understand the effects of our key idea, applying mutations at different levels of abstraction, we conduct an ablation study. This follows best practices~\cite{schloegel2024sok} and serves to measure individual design decisions in isolation. This study is designed to isolate the contribution of mutations on individual layers, showing their contribution to branch coverage. To this end, we test four ablations of \toolname over a 48-hour period, running each ablation ten times. None of the configurations have access to informed seeds. Instead, the initial corpus is composed of samples produced by our generator, as described in~\Cref{sec:design-generator}. After generating the initial corpus, samples are mutated as also described in~\Cref{sec:design-mutator}.

\vspace{0.2em}
\begin{itemize}
\item \textbf{IR disabled}, a configuration with IR mutations disabled. All mutations are solely based on tree-operations.
\item \textbf{AST disabled}, a variant that disables AST mutations and performs its mutations only on the IR layer.
\item \textbf{IR delayed}, a variant of \toolname performing exclusively AST mutations during the first 24h. After 24h, we enable IR mutations as well. The purpose of this ablation is to measure whether IR mutations contribute additional coverage on a corpus constructed from AST mutations.
\item \textbf{AST delayed}, a variant performing exclusively IR mutations during the first 24h. Similar to \textsc{IR~delayed} but with roles reversed (\ie we enable AST mutations after the 24h mark), we measure whether AST mutations contribute additional coverage on a corpus constructed from IR mutations.
\end{itemize}

The results of the ablation experiments on \tint, \wgslc, and \dxc are depicted in~\Cref{fig:ablation}, the results on \naga in \Cref{fig:naga_ablation}. Our analysis shows that on the front-end translators \tint, \wgslc, and \naga, ablations that prioritize AST mutations (\textsc{IRdelayed} and \textsc{IRdisabled}) perform better than those focusing on IR mutations. Notably, introducing IR mutations after 24 hours in the \textsc{IRdelayed} setup does not improve coverage compared to the \textsc{IRdisabled} scenario. This observation aligns with the fact that shader front-ends primarily handle parsing and transforming the parsed AST, suggesting that mutations at this level thoroughly cover the SUT.

Conversely, the back-end compiler \dxc shows different results. Here, configurations with IR mutations outperform those with AST mutations. Enabling AST mutations after 24 hours does not enhance branch coverage. \dxc is based on LLVM~\cite{lattner2004llvm} and its primary purpose is optimizing code with a pipeline of optimization passes. Each pass transforms the LLVM IR, with the goal of producing faster code. The fact that IR mutations are more effective for testing \dxc corroborates findings from front-end translators: Mutations operating at a similar abstraction layer as the SUT are the most effective ones.
We believe this to be the consequence of these mutations having a direct impact on the \emph{form} of the input as considered by the target, while mutations on another abstraction level may have no semantic impact on this form of the input. For example, replacing one variable by another on the IR level may change the whole meaning of the IR, but the AST's structure remains identical. 

\begin{takeaway}
AST mutations work best for exploring the front-end, while IR mutations excel at covering branches in the back-end. Testing the entire pipeline requires a combination of both.
\end{takeaway}

The \textsc{IRdelayed} ablation leads to a noteworthy observation when evaluating coverage in the latter half of the experiment, specifically after activating IR mutations on a corpus initially created through AST mutations. It is important to recall that the fuzzing process for \dxc first sends \wgsl shaders through \tint for translation into HLSL before passing them to \dxc. Enabling IR mutations does not affect \tint's coverage, but when the HLSL output from \tint is processed by \dxc, there is a noticeable improvement in coverage.
This observation implies that delayed IR mutations do not influence the front-end translator \tint but significantly impact the downstream component \dxc. \wgsl shaders contain inherent complexities that AST mutations alone cannot adequately explore, confirming our design strategy, which integrates both AST and IR mutations.

\begin{takeaway}
Operations that have no effect in the front-end application may still have a large impact on the back-end.
\end{takeaway}

\begin{table*}[tb]
    \centering
    \caption{Overview of the \numbugs bugs we found in different targets. All bugs have been responsibly disclosed and reported pseudonymously (as `wgslfuzz'). \new{The column \emph{Bug ID}} links to the associated CVE record or bug report. At the time of submission, not all bug reports have been made public by the respective maintainers due to security concerns.}%
    \label{tab:Bugs}

    \begin{tabular}{ l l l l l }
        \toprule
        \textbf{SUT} & \textbf{Bug ID} & \textbf{Browser} & \textbf{Status} & \textbf{Description} \\
        \midrule
        angle & \bugreportchromium{329271490} & \chromelogo\hfill\firefoxlogo\hfill\safarilogo & fixed & Stack out-of-bound access in shader translation \\
        dxcompiler & \bugreportchromium{1513069} & \chromelogo & open & Heap OOB in \texttt{dxil} writer due to large binding ids \\
        dxcompiler & \bugreportcve{CVE-2024-2885} & \chromelogo & fixed & Heap UAF in dxcompiler via tint generated shader \\ %
        dxcompiler & \bugreportcve{CVE-2024-3515} & \chromelogo & fixed & Heap UAF in dxcompiler via tint generated shader \\ %
        dxcompiler & \bugreportcve{CVE-2024-4948} & \chromelogo & fixed & Heap UAF in dxcompiler via tint generated shader \\
        dxcompiler & \bugreportcve{CVE-2024-4060} & \chromelogo & fixed & UAF in dxcompiler via tint generated shader \\
        dxcompiler & \bugreportcve{CVE-2024-4368} & \chromelogo & fixed & Memory safety violation in dxcompiler via tint generated shader \\
        \new{dxcompiler} & \bugreportcve{CVE-2024-5160} & \new{\chromelogo} & \new{fixed} & \new{Heap OOB via tint generated shader} \\
        \new{dxcompiler} & \bugreportcve{CVE-2024-5494} & \new{\chromelogo} & \new{fixed} & \new{Heap UAF due to incorrect removal of \texttt{switch} statements}  \\
        \new{dxcompiler} & \bugreportcve{CVE-2024-5495} & \new{\chromelogo} & \new{fixed} & \new{Heap UAF due to incorrect removal of phi nodes} \\
        \new{dxcompiler} & \bugreportcve{CVE-2024-6102} & \new{\chromelogo} & \new{fixed} & \new{Heap OOB due to broken control flow} \\
        \new{dxcompiler} & \bugreportcve{CVE-2024-5831} & \new{\chromelogo} & \new{fixed} & \new{Heap UAF caused by incorrect dead-code elimination} \\
        \new{dxcompiler} & \bugreportcve{CVE-2024-5832} & \new{\chromelogo} & \new{fixed} & \new{Heap UAF due to incorrect phi node update} \\
        \new{dxcompiler} & \bugreportcve{CVE-2024-6290} & \new{\chromelogo} & \new{fixed} & \new{Heap UAF caused by incorrect vector flattening} \\
        \new{dxcompiler} & \bugreportcve{CVE-2024-6292} & \new{\chromelogo} & \new{fixed} & \new{Heap UAF due to incorrect instruction folding}  \\
        \new{dxcompiler} & \bugreportcve{CVE-2024-6103} & \new{\chromelogo} & \new{fixed} & \new{Heap UAF when replacing phi nodes with select instructions} \\
        \new{dxcompiler} & \bugreportcve{CVE-2024-6293} & \new{\chromelogo} & \new{fixed} & \new{Heap UAF caused by incorrect loop induction optimization} \\
        \new{dxcompiler} & \bugreportcve{CVE-2024-6991}     & \new{\chromelogo} & \new{fixed}  & \new{Stack use-after-return during lowering of matrix instructions} \\
        \new{tint} & \bugreporttint{2190}                    & \new{\chromelogo} & \new{fixed} & \new{ICE: Error during type validation results in crash}  \\
        \new{tint} & \bugreporttint{2201}                    & \new{\chromelogo} & \new{fixed} & \new{ICE: Reached an \textit{unreachable()}, in turn crashing the SUT} \\
        \new{tint} & \bugreporttint{2202}                    & \new{\chromelogo} & \new{fixed} & \new{Near-null deref in IR shader translator} \\
        tint & \bugreporttint{2055} & \chromelogo & fixed & ICE: Incorrect validation of pointers-to-pointers \\
        tint & \bugreporttint{2056} & \chromelogo & fixed & ICE: Incorrect typing of \texttt{array()} with mixed types \\
        tint & \bugreporttint{2058} & \chromelogo & fixed & ICE: Incomplete types used as sub-types trigger a crash \\
        tint & \bugreporttint{2068} & \chromelogo & fixed & Accepting a malformed shader triggered an ICE \\
        tint & \bugreporttint{2076} & \chromelogo & fixed & ICE: crash when multiple entry points duplicate bindings \\
        tint & \bugreporttint{2077} & \chromelogo & fixed & ICE: \texttt{MergeReturn()} crashed when emitting an exit instruction \\
        tint & \bugreporttint{2078} & \chromelogo & fixed & SPIR-V validation: Missing constructor calls \\
        tint & \bugreporttint{2079} & \chromelogo & fixed & SPIR-V validation: Incorrect vector code generation \\
        tint & \bugreporttint{2092} & \chromelogo & open & Error in the SPIR-V validator itself \\
        tint & \bugreporttint{2194} & \chromelogo & open & SPIR-V validation: Invalid codegen for \texttt{OpConstantComposite} \\
        naga & \bugreportnaga{2560} & \multicolumn{1}{c}{\firefoxlogo} & fixed & OOM triggered when compiling wgsl shader \\
        naga & \bugreportnaga{2568} & \multicolumn{1}{c}{\firefoxlogo} & fixed & Index out of bounds in expression lowering \\
        naga & \bugreportwgpu{4547} & \multicolumn{1}{c}{\firefoxlogo} & open & Index out of bounds in analyzer \\
        naga & \bugreportwgpu{4512} & \multicolumn{1}{c}{\firefoxlogo} & open & Internal error: entered unreachable code \\
        naga & \bugreportwgpu{4513} & \multicolumn{1}{c}{\firefoxlogo} & open & Panic in HLSL writer when translating push constants \\
        naga & \bugreportwgpu{5547} & \multicolumn{1}{c}{\firefoxlogo} & fixed & Accepting a malformed shader results in invalid SPIR-V code \\
        wgslc & \bugreportwebkit{268148} & \multicolumn{1}{r}{\safarilogo} & open & Heap UAF in \texttt{invalidateIterators} \\
        wgslc & \bugreportwebkit{273407} & \multicolumn{1}{r}{\safarilogo} & fixed & Assertion violation during type inference \\
        wgslc & \bugreportwebkit{273411} & \multicolumn{1}{r} {\safarilogo} & fixed & Type checker asserts during parsing of corrupted shader \\
        \bottomrule
    \end{tabular}
\end{table*}

\subsection{Found Bugs}
Excavating new and interesting bugs is the key feature of any fuzzer. To test the effectiveness of our tool, we ran multiple fuzzing campaigns with \toolname on \wgslc, \tint, \dxc, and \naga. It is noteworthy that the latter is written in memory-safe Rust. Hence, out-of-bound accesses and other issues traditionally plaguing C/C++ are not a security concern and affect availability only. However, logical flaws resulting in mistranslated shaders affect even memory-safe languages.
Over the course of several weeks, we discovered \numbugs bugs in total, with at least one bug in each scrutinized component.
Despite the shader translators being tested by their respective vendors with fuzzing (\eg \bugreportchromium{335245351}), \toolname uncovered a wide variety of bugs in various stages of the shader compilation pipeline.
Bug classes identified during fuzzing include out-of-bound-accesses in the translator front-end, incorrect code emission in the translator back-end, and memory-safety violations in OS-specific compilers. 
We provide a complete list of all issues identified so far in~\Cref{tab:Bugs}.

\paragraph{Case Study: Incorrect Code Generation} When fuzzing \wgsl shader translators, we discovered several logical errors during shader translation. Specific instances include errors \emph{\bugreportwgpu{5547}} and \emph{\bugreporttint{2079}}. In both cases, the respective shader translator emitted incorrect SPIR-V code, potentially causing subsequent errors in downstream shader compilers. It is important to note that this type of error also impacts \naga. Although \naga is safeguarded from memory safety violations due to its use of the Rust programming language, it is still susceptible to generating faulty code from logical errors, a problem common across all programming languages. The issue in \naga is particularly notable because the root cause is not an error in the SPIR-V specific back-end of the shader translator. Instead, the flaw stems from the \wgsl validation pass, which \emph{should} reject invalid shaders. Due to an oversight, the validator accepts a malformed \wgsl shader, leading to incorrect SPIR-V code. This oversight is problematic in the context of SPIR-V because if such a flawed shader is accepted, it puts other \naga lifters at risk, such as HLSL and Metal, none of which can handle malformed shaders. 

\paragraph{Case Study: Memory Safety Violations} In addition to logic bugs, our fuzzing campaigns on shader translators uncovered multiple memory-safety violations.
Note that we also found out-of-bound accesses in \naga. However, these are strongly mitigated by the Rust programming language.
In the following, we put the spotlight on two bugs in \dxc, emphasizing the need to test not only the front-end translators but also downstream components.
In particular, the effectiveness of this fuzzing pipeline hinges on the semantic correctness of the generated and mutated shaders, as only semantically correct shaders are passed to back-end compilers.

One of the security vulnerabilities found by \toolname is CVE-2024-3515. \Cref{lst:multidelete} shows an HLSL shader that triggered a memory corruption in affected versions of \dxc (this was fixed after our reporting). Specifically, in line 7 of the HLSL shader, there is a self-assignment of a static struct. During optimization, this self-assignment is correctly identified as redundant and marked for removal. However, in the function \texttt{ScalarReplAggregatesHLSL}, both the source and target of the self-assignment are deleted separately. Failing to handle the corner case where target and source of the assignment are identical leads to a double-delete error in the Chrome GPU process.

Another HLSL shader triggers a memory corruption in \dxc as shown in~\Cref{lst:dceuaf}. The memory corruption in the Chrome GPU process has been assigned CVE-2024-2885. The root cause of the issue is an HLSL-specific optimization pass in \dxc called \texttt{HLMatrixLowerPass}. At the beginning of the optimization, the pass first extends the LLVM IR with an internal stub function. Later, a call to this stub is added. The call instruction is added between the IR code corresponding to lines 3 and 5.
Once the pass finishes, the stub function is deleted. However, the inserted call instruction remains. In consequence, the call instruction references the deleted stub function. Later, a dead-code-elimination pass correctly identifies all code following the while loop (line 2) as dead, because the loop contains no break condition.
Attempting to remove the dead code containing a dangling pointer ultimately results in a memory-safety violation, crashing the GPU process.

\begin{figure}[tb]
\begin{subfigure}[b]{.47\linewidth}
\centering
\begin{minted}[%
    frame=lines,
    framesep=2mm,
    fontsize=\footnotesize,
    stripnl=false,
    linenos,
    xleftmargin=1.8em
]{hlsl}
struct MyStruct {
  int m0;
};
static MyStruct s;

void foo() {
  s = s; // dead assignment
}

[numthreads(1, 1, 1)]
void main() {
  foo();
}
\end{minted}
\caption{CVE-2024-3515}
\label{lst:multidelete}
\end{subfigure}
\hfill
\begin{subfigure}[b]{.47\linewidth}
\centering
\begin{minted}[%
    frame=lines,
    framesep=2mm,
    fontsize=\footnotesize,
    stripnl=false,
    linenos,
    xleftmargin=1.8em
]{hlsl}
float4x2 foo() {
  while (true) {
  }
  // intrinsic call inserted
  return float4x2(
    (0.0f).xx, (0.0f).xx,
    (0.0f).xx, (0.0f).xx);
}

[numthreads(1, 1, 1)]
void main() {
  float4x2 e = foo();
}
\end{minted}
\caption{CVE-2024-2885}
\label{lst:dceuaf}
\end{subfigure}
\caption{HLSL shaders generated by \tint that trigger memory safety violations in \dxc, a component of the Chrome GPU process. (a) Optimizing the self-assignment of the static struct in line 7 leads to a double-free. (b) The infinite loop in line 2 invokes dead-code elimination, which operates on freed IR objects.}
\end{figure}

\paragraph{\new{Real-world Impact}}
\new{Our whole testing setup is designed to identify bugs that are relevant in practice, \ie that can be triggered by attacker-controlled input. From an attacker's point of view, triggering this bug is as simple as embedding the fuzzer output (i.e., a shader) into an HTML file and serving it to unsuspecting victims. Any user visiting this website automatically processes this shader via their browsers, triggering bugs in affected components. This makes memory corruption bugs in the shader pipeline so security-sensitive, as confirmed by the vendors. Even worse, Firefox does not sandbox this highly privileged process, and Chrome only uses a weaker sandbox than for other web content. %
For exemplary HTML files containing malicious shaders, we refer to the bugs reported in \Cref{tab:Bugs} that have been assigned a CVE.
Technical details of all bugs can be found in the issue tracker linked in the CVE entry. 
}

\section{Discussion}%
\label{sec:Discussion}
The proposed approach of mutating shaders on both an IR layer and an AST layer is suitable for fuzzing the WebGPU pipeline end-to-end, \ie from the initial parsing in the browser deep into the transformation passes of back-end compilers.
Our approach was successful in uncovering a multitude of bugs and performs favorably in other evaluation metrics. In the following section, we discuss shortcomings of \toolname and potential future work.

\paragraph{Threats to Validity}
Ensuring the accuracy of conclusions from empirical experiments is essential. We focus on three key aspects to validate our findings and describe assumptions and methodologies:

\emph{External Validity.}
One vital concern is whether the results from our tested programs are transferable to other related targets, such as compilers in GPU drivers and the Mesa project. While predicting results for untested software is difficult, we assessed \toolname across all three available translators from \wgsl to OS-specific back-ends and the back-end compiler \dxc. To enhance the validity of our approach, we will release our implementation under an open-source license, allowing others to test and evaluate it.

\emph{Internal Validity.}
To improve the accuracy of our evaluation, we conducted each experiment ten times across all targets to minimize systematic errors. Additionally, we measured branch coverage for all fuzzing campaigns using the same lcov-instrumented binary, ensuring consistent coverage data. However, the results from fuzzers having access to an informed corpus lack comparability to fuzzers without such seeds. For example, outcomes from \toolname cannot be directly compared with those from \wgslgenerator. To address this, we included an uninformed version of our approach, \toolnameNoSeeds, in our experiments. Lastly, we used the same seed files for all informed fuzzers to maintain comparable results.

\emph{Construct Validity.}
A primary concern about validity is ensuring that an evaluation truly measures what it is intended to. Directly comparing the \emph{concepts} of different fuzzers is impossible, as we can only evaluate tools that embody their respective design.
This makes a fair evaluation of concepts challenging, because outcomes are heavily affected by unrelated elements, like algorithmic optimizations and fine-tuned parameters~\cite{rizzi2016techniques}. For example, the throughput of
\wgslgenerator is less than $1$ samples per second, compared to more than $100$ for \regexfuzzer. This difference influences the branch coverage achieved on a SUT, potentially biasing evaluation results towards performance-optimized fuzzers.
By examining additional metrics like the semantic correctness rate, which is independent of throughput, we provide a less distorted picture.
Furthermore, we performed an ablation study of our implementation that allows the attribution of differences in coverage to contributing design factors, as opposed to mere implementation details.

\paragraph{Seeds}
Our method interleaves shader generation and mutations and is the first \wgsl fuzzer reaching high branch coverage without pre-existing informed seeds. Still, our evaluation in~\Cref{sec:eval-coverage} shows that using informed seeds enhances our tool's ability to explore target applications. Improving branch coverage with an informed seed corpus suggests that our technique for generating and mutating samples has room for improvement.
We can identify where enhancements could significantly increase coverage by analyzing the coverage differences between \toolnameNoSeeds and the informed seed corpus.

\paragraph{Back-ends and GPU Drivers}
We are actively testing the translation capabilities of all three major browsers and the DirectX system, the latter being specific to Microsoft Windows. However, there is still a need for exploring additional systems and compilation phases. For instance, the Mesa 3D Graphics Library on Linux is responsible for converting SPIR-V code generated by browsers into a format specific to the available GPU. Mesa does not expose an interface suitable for fuzzing SPIR-V shader translation;  implementing such an interface would enable testing of a wider range of back-end shader compilers. On macOS, the back-end component compiling Metal shaders (analog to \dxc for HLSL) is not open-source. Therefore, neither our coverage instrumentation nor ASAN instrumentation is directly applicable. We leave the integration of the component with a binary-only fuzzer as future work.

Our testing efforts are focused solely on components outside the kernel, \ie userland components. Nonetheless, a portion of the GPU processing occurs within the kernel. For example, DirectX communicates with the kernel using a version of LLVM~3.7 bitcode, which the GPU's device driver compiles further to an internal ISA. Hardening the interface between untrusted shaders passed from user-mode to privileged components via fuzzing improves overall system security. Resetting device drivers and the operating system in between fuzzing inputs poses a significant challenge, to which snapshot fuzzing~\cite{schumilo2021nyx} presents a viable solution.

\paragraph{Differential Testing}
The front-end translators \naga, \wgslc, and \tint all adhere to the WebGPU Shader Language specification. As a result, they are designed to consistently accept a uniform set of valid input shaders and reject invalid ones. Should discrepancies arise among these shader translators, they often indicate a misinterpretation of the \wgsl specification by one of the translators or a potential ambiguity within the specification itself.

The methodology for differential testing of shader compilers offers further room for advancement. 
A single compute shader should yield identical computational results across all implementations once executed by OS-specific back-ends. This approach ensures a high level of consistency in shader execution, which benefits the development of cross-platform graphics applications.

\paragraph{Retrofitting Memory Safety} The back-end compiler \dxc optimizes shader programs with complex analyses and code transformations. This complexity, along with the fact that the component has not been sufficiently hardened against adversarial inputs, continues to pose a security threat. Converting the C/C++ code to WebAssembly~\cite{narayan2020retrofitting,haas2017bringing} could allow for fine-grained sandboxing of \dxc, reducing the impact of bugs.

\section{Related Work}%
\label{sec:RelatedWork}

In this work, we have introduced \toolname, a fuzzer highly effective in uncovering security bugs in shaders. However, we are not the first to stress-test WebGPU or the graphics stack. Tools similar to our work include \regexfuzzer~\cite{donaldson2023industrial} and \astfuzzer~\cite{donaldson2023industrial}, two coverage-guided fuzzers for \wgsl. Both of these tools are tailored specifically to \tint and do not support other SUTs. The performance of these two approaches heavily relies on the quality of their seed corpus, in particular these tool \emph{require} an informed corpus. In contrast to \toolname, neither of these two tools includes mutations on an IR layer. As a result, their ability to mutate aggregate data types, control flow, and complex statements is either limited in scope or completely absent.

\wgslsmith~\cite{mohsin2022wgslsmith} and \wgslgenerator~\cite{wgslgenerator} are domain-specific tools to generate \wgsl code. These tools draw inspiration from Csmith~\cite{yang2011finding}, a well-known method also adapted for testing other graphics components such as CUDA~\cite{jiang2020cudasmith}. Unlike approaches involving mutations, \wgslsmith and \wgslgenerator rely solely on code generation. This implies that they do not utilize a pre-existing seed corpus to improve output quality. As demonstrated in \Cref{fig:coverage}, this generational approach results in less comprehensive branch coverage. Furthermore, as \Cref{tab:semRate} illustrates, most of the code samples produced by these tools are ultimately rejected by their intended targets.

Outside the scope of WebGPU, GraphicsFuzz~\cite{graphicsfuzz}, including its components GL-Fuzz~\cite{donaldson2017automated} and spirv-fuzz~\cite{donaldson2020putting}, is designed for metamorphic testing to identify rendering bugs in graphic shader compilers. Unlike \toolname, which targets memory safety violations in \wgsl shaders, GraphicsFuzz specifically examines inconsistencies across different metamorphic variants~\cite{chen1998metamorphic, segura2016survey, le2014compiler} of SPIR-V and GLSL shaders.

In an even broader perspective, grammar fuzzing~\cite{aschermann2019nautilus, srivastava2021gramatron} is used to find bugs by generating structured inputs based on grammar rules. It involves creating test cases that adhere to the syntax and semantics of the targeted system, often using context-free grammars. While conceptually applicable to \wgsl, there are currently no grammar fuzzers specifically targeting \wgsl. Another interesting approach to test the shader compilation pipeline could involve differential testing~\cite{bernhard2022jitpicker,graphicsfuzz}.

Lastly, the security of the graphics stack is broader than \wgsl shader compilation. Web-exposed APIs for resource management have been tested in the context of WebGL~\cite{peng2023gleefuzz} and the entire scope of the browser~\cite{dominowebidlfuzzer,dinh2021favocado}. \new{Even broader, related work has tested other parts of the browser, including the DOM~\cite{xu2020freedom,zhou2023bettersemantics,zhou2022minerva} or security policies~\cite{shou2021corbfuzz,kim2022fuzzorigin}.} While all parts of the browser deserve close scrutiny, bugs found in these components are often mitigated by browser hardening mechanisms, such as the sandbox.

\section{Conclusion}%
\label{sec:Conclusion}
In this paper, we presented the design and implementation of \toolname, a comprehensive fuzzing framework for the WebGPU shader language. We proposed the idea of fuzzing WGSL on two abstraction levels, namely on an IR layer and an AST layer.
The mutations on both abstraction layers complement each other, as each of them can transform inputs in a manner the other one cannot. For example, while the IR layer is suitable for mutating and generating additional types, it cannot produce ASTs that violate the specification.
On the other hand, the AST layer allows transformations that stress the parser. Consequently, the synergy of the two mutation layers allows thorough testing of target applications.
In our evaluation, \toolname outperforms domain-specific fuzzers in terms of code coverage, as well as performing favorably in terms of semantic correctness rate. Furthermore, our approach shows its real-world impact by identifying \numbugs bugs in components exposed via the Internet, with \numcves CVEs assigned so far.

\begin{acks}
We thank our reviewers for their valuable feedback, and Florian Bauckholt, Tobias Scharnowski, and Sahil Sihag for their thoughts on a draft of this work.
We would like to acknowledge Google's and Mozilla's swift and decisive response to our bug reports.
This work was funded by the European Research Council (ERC) under the consolidator grant RS$^3$ (101045669). %
\end{acks}

\appendix

\newpage

\onecolumn

\section{\naga Coverage and Ablation Study}

\begin{figure*}[h]
\centering
\begin{subfigure}[b]{0.4\linewidth}
    \centering
    \graphicspath{{figures/}}
    \def\svgwidth{\textwidth}
    \begin{footnotesize}
        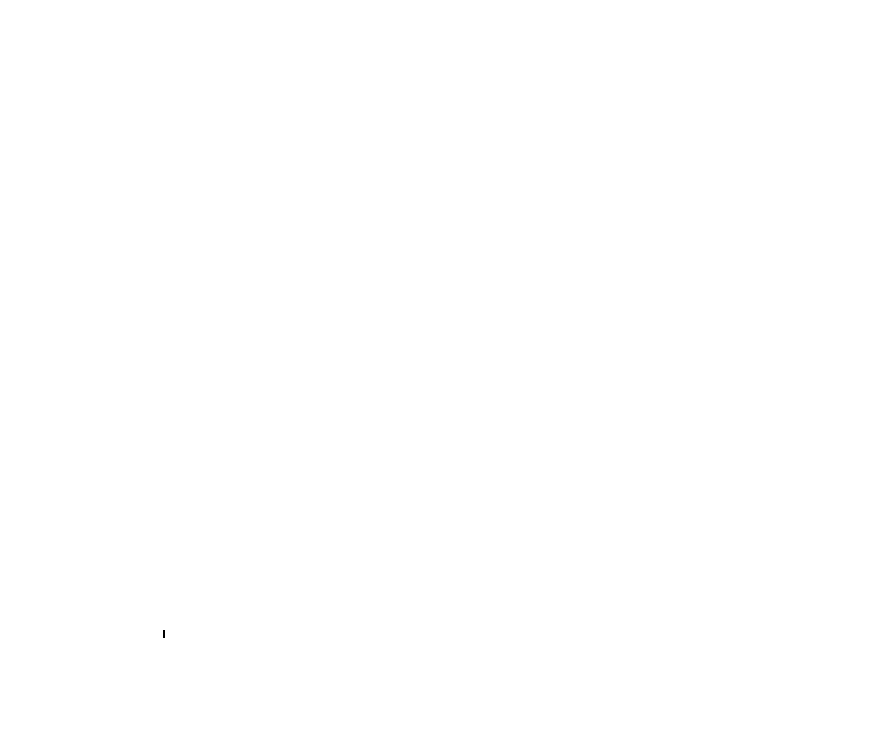
    \end{footnotesize}
    \caption{\naga line coverage over 24h.}%
    \label{fig:naga_coverage}
\end{subfigure}
\hspace{4em}
\begin{subfigure}[b]{0.4\linewidth}
    \centering
    \graphicspath{{figures/}}
    \def\svgwidth{\textwidth}
    \begin{footnotesize}
        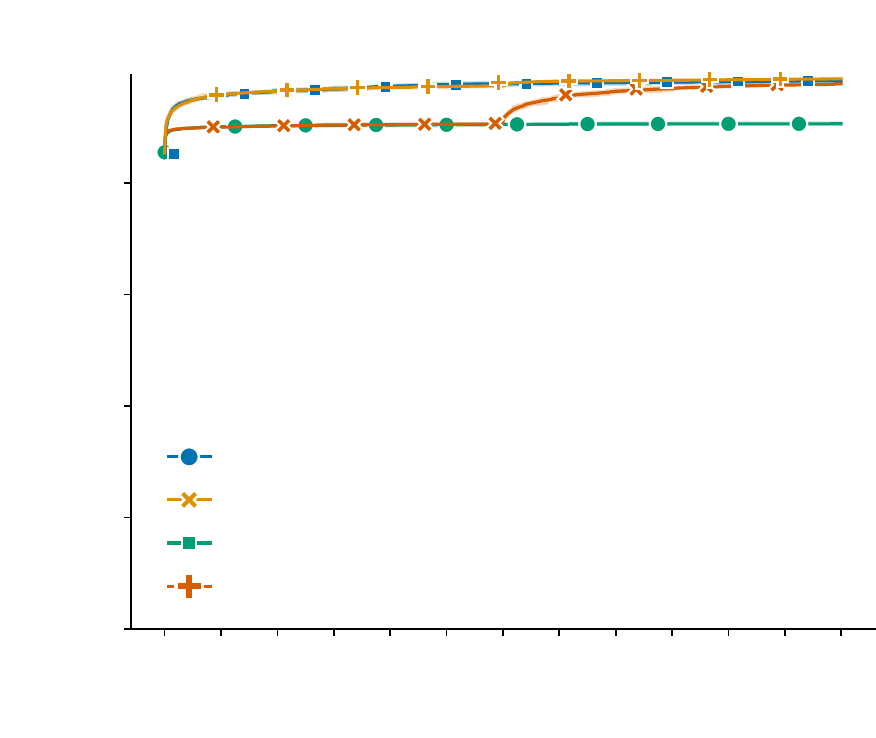
    \end{footnotesize}
    \caption{\naga line coverage ablation study.}%
    \label{fig:naga_ablation}
\end{subfigure}
\caption{We measure line coverage for \naga, as it is Rust-based, making branch coverage a suboptimal metric. For a more in-depth explanation for using line coverage over branch coverage, refer to~\Cref{sec:eval-coverage}.}
\end{figure*}


\begin{thebibliography}{10}

\bibitem{urlWebGPUfirefox}
{ Dzmitry Malyshau}.
\newblock {A Taste of WebGPU in Firefox}.
\newblock
  \url{https://hacks.mozilla.org/2020/04/experimental-webgpu-in-firefox/},
  2020.

\bibitem{tintGoogle}
{ googlesource.com }.
\newblock {Tint is a compiler for the WebGPU Shader Language (WGSL)}.
\newblock \url{https://dawn.googlesource.com/tint}, 2024.

\bibitem{wgslgenerator}
{ Hana Watson }.
\newblock {WGSLGenerator}.
\newblock \url{https://github.com/hanawatson/wgslgenerator}, 2022.

\bibitem{urlHSLSMicrosoft}
{ Microsoft }.
\newblock {High-level shader language (HLSL)}.
\newblock
  \url{https://learn.microsoft.com/en-us/windows/win32/direct3dhlsl/dx-graphics-hlsl},
  2021.

\bibitem{nagalib}
{ The naga authors }.
\newblock {naga}.
\newblock \url{https://github.com/gfx-rs/wgpu/tree/trunk/naga}, 2023.

\bibitem{urlMetalShadingLang}
{Apple Inc.}
\newblock {Metal Shading Language Specification}.
\newblock
  \url{https://developer.apple.com/metal/Metal-Shading-Language-Specification.pdf},
  2023.

\bibitem{aschermann2019nautilus}
Cornelius Aschermann, Tommaso Frassetto, Thorsten Holz, Patrick Jauernig,
  Ahmad-Reza Sadeghi, and Daniel Teuchert.
\newblock {Nautilus: Fishing for Deep Bugs with Grammars}.
\newblock In {\em Symposium on Network and Distributed System Security (NDSS)},
  2019.

\bibitem{aschermann2019redqueen}
Cornelius Aschermann, Sergej Schumilo, Tim Blazytko, Robert Gawlik, and
  Thorsten Holz.
\newblock {RedQueen: Fuzzing with Input-to-State Correspondence}.
\newblock In {\em Symposium on Network and Distributed System Security (NDSS)},
  2019.

\bibitem{bars2023fuzztruction}
Nils Bars, Moritz Schloegel, Tobias Scharnowski, Nico Schiller, and Thorsten
  Holz.
\newblock {Fuzztruction: Using Fault Injection-based Fuzzing to Leverage
  Implicit Domain Knowledge}.
\newblock In {\em USENIX Security Symposium}, 2023.

\bibitem{bernhard2022jitpicker}
Lukas Bernhard, Tobias Scharnowski, Moritz Schloegel, Tim Blazytko, and
  Thorsten Holz.
\newblock {JIT-Picking: Differential Fuzzing of JavaScript Engines}.
\newblock In {\em ACM Conference on Computer and Communications Security
  (CCS)}, 2022.

\bibitem{bohme2016coverage}
Marcel B{\"o}hme, Van-Thuan Pham, and Abhik Roychoudhury.
\newblock {Coverage-based Greybox Fuzzing as Markov Chain}.
\newblock In {\em ACM Conference on Computer and Communications Security
  (CCS)}, 2016.

\bibitem{chen1998metamorphic}
Tsong~Y Chen, Shing~C Cheung, and Shiu~Ming Yiu.
\newblock {Metamorphic Testing: A New Approach for Generating Next Test Cases}.
\newblock Technical report, Hong Kong University of Science and Technology,
  1998.

\bibitem{chen2021one}
Yongheng Chen, Rui Zhong, Hong Hu, Hangfan Zhang, Yupeng Yang, Dinghao Wu, and
  Wenke Lee.
\newblock {One Engine to Fuzz’em All: Generic Language Processor Testing with
  Semantic Validation}.
\newblock In {\em IEEE Symposium on Security and Privacy (S\&P)}, 2021.

\bibitem{urlChromeSandbox}
{Chromium}.
\newblock {Chromium Docs: Sandbox}.
\newblock
  \url{https://chromium.googlesource.com/chromium/src/+/HEAD/docs/design/sandbox.md},
  2024.

\bibitem{urlGPUwebExplainer}
{Contributors to the WebGPU Explainer Specification}.
\newblock {WebGPU Explainer}.
\newblock \url{https://gpuweb.github.io/gpuweb/explainer/}, 2024.

\bibitem{dinh2021favocado}
Sung~Ta Dinh, Haehyun Cho, Kyle Martin, Adam Oest, Kyle Zeng, Alexandros
  Kapravelos, Gail-Joon Ahn, Tiffany Bao, Ruoyu Wang, Adam Doup{\'e}, et~al.
\newblock {Favocado: Fuzzing the Binding Code of JavaScript Engines Using
  Semantically Correct Test Cases}.
\newblock In {\em Symposium on Network and Distributed System Security (NDSS)},
  2021.

\bibitem{donaldson2023industrial}
Alastair~F Donaldson, Ben Clayton, Ryan Harrison, Hasan Mohsin, David Neto,
  Vasyl Teliman, and Hana Watson.
\newblock {Industrial Deployment of Compiler Fuzzing Techniques for Two GPU
  Shading Languages}.
\newblock In {\em IEEE Conference on Software Testing, Verification and
  Validation (ICST)}, 2023.

\bibitem{donaldson2017automated}
Alastair~F Donaldson, Hugues Evrard, Andrei Lascu, and Paul Thomson.
\newblock {Automated Testing of Graphics Shader Compilers}.
\newblock {\em Proceedings of the ACM on Programming Languages (PACMPL)}, 2017.

\bibitem{donaldson2020putting}
Alastair~F Donaldson, Hugues Evrard, and Paul Thomson.
\newblock {Putting Randomized Compiler Testing into Production (Experience
  Report)}.
\newblock In {\em European Conference on Object-Oriented Programming (ECOOP)},
  2020.

\bibitem{fioraldi2022libafl}
Andrea Fioraldi, Dominik~Christian Maier, Dongjia Zhang, and Davide Balzarotti.
\newblock {LibAFL: A Framework to Build Modular and Reusable Fuzzers}.
\newblock In {\em ACM Conference on Computer and Communications Security
  (CCS)}, 2022.

\bibitem{gross2023fuzzilli}
Samuel Gro{\ss}, Simon Koch, Lukas Bernhard, Thorsten Holz, and Martin Johns.
\newblock {Fuzzilli: Fuzzing for JavaScript JIT Compiler Vulnerabilities}.
\newblock In {\em Symposium on Network and Distributed System Security (NDSS)},
  2023.

\bibitem{haas2017bringing}
Andreas Haas, Andreas Rossberg, Derek~L Schuff, Ben~L Titzer, Michael Holman,
  Dan Gohman, Luke Wagner, Alon Zakai, and JF~Bastien.
\newblock {Bringing the Web up to Speed with WebAssembly}.
\newblock In {\em ACM SIGPLAN Conference on Programming Language Design and
  Implementation (PLDI)}, 2017.

\bibitem{han2019codealchemist}
HyungSeok Han, DongHyeon Oh, and Sang~Kil Cha.
\newblock {CodeAlchemist: Semantics-aware Code Generation to Find
  Vulnerabilities in JavaScript Engines}.
\newblock In {\em Symposium on Network and Distributed System Security (NDSS)},
  2019.

\bibitem{Hatch23Generating}
William~Gallard Hatch, Pierce Darragh, Sorawee Porncharoenwase, Guy Watson, and
  Eric Eide.
\newblock {Generating Conforming Programs with Xsmith}.
\newblock In {\em ACM SIGPLAN International Conference on Generative
  Programming: Concepts \& Experiences (GPCE)}, 2023.

\bibitem{holler2012fuzzing}
Christian Holler, Kim Herzig, and Andreas Zeller.
\newblock {Fuzzing with Code Fragments}.
\newblock In {\em USENIX Security Symposium}, 2012.

\bibitem{dominowebidlfuzzer}
{Jason Kratzer}.
\newblock {Fuzzing Firefox with WebIDL}.
\newblock \url{https://hacks.mozilla.org/2020/04/fuzzing-with-webidl/}, 2020.

\bibitem{jiang2020cudasmith}
Bo~Jiang, Xiaoyan Wang, Wing~Kwong Chan, TH~Tse, Na~Li, Yongfeng Yin, and
  Zhenyu Zhang.
\newblock {Cudasmith: A Fuzzer for Cuda Compilers}.
\newblock In {\em IEEE Annual Computers, Software, and Applications Conference
  (COMPSAC)}, 2020.

\bibitem{kim2022fuzzorigin}
Sunwoo Kim, Young~Min Kim, Jaewon Hur, Suhwan Song, Gwangmu Lee, and
  Byoungyoung Lee.
\newblock {FuzzOrigin: Detecting UXSS vulnerabilities in Browsers through
  Origin Fuzzing}.
\newblock In {\em USENIX Security Symposium}, 2022.

\bibitem{klees2018evaluating}
George Klees, Andrew Ruef, Benji Cooper, Shiyi Wei, and Michael Hicks.
\newblock {Evaluating Fuzz Testing}.
\newblock In {\em ACM Conference on Computer and Communications Security
  (CCS)}, 2018.

\bibitem{lattner2004llvm}
Chris Lattner and Vikram Adve.
\newblock {LLVM: A Compilation Framework for Lifelong Program Analysis \&
  Transformation}.
\newblock In {\em International Symposium on Code Generation and Optimization
  (CGO)}, 2004.

\bibitem{le2014compiler}
Vu~Le, Mehrdad Afshari, and Zhendong Su.
\newblock {Compiler Validation via Equivalence Modulo Inputs}.
\newblock {\em ACM SIGPLAN Notices}, 2014.

\bibitem{llvmcov}
{LLVM Project}.
\newblock {llvm-cov -- Emit Coverage Information}.
\newblock \url{https://llvm.org/docs/CommandGuide/llvm-cov.html}.

\bibitem{lyu2019mopt}
Chenyang Lyu, Shouling Ji, Chao Zhang, Yuwei Li, Wei-Han Lee, Yu~Song, and
  Raheem Beyah.
\newblock {MOPT: Optimized Mutation Scheduling for Fuzzers}.
\newblock In {\em USENIX Security Symposium}, 2019.

\bibitem{treesitter}
{Max Brunsfeld}.
\newblock {tree-sitter}.
\newblock \url{https://tree-sitter.github.io/tree-sitter/}, 2018.

\bibitem{mohsin2022wgslsmith}
Hasan Mohsin.
\newblock {WGSLsmith: A Random Generator of WebGPU Shader Programs}.
\newblock Master's thesis, Imperial College London, 2022.

\bibitem{urlFirefoxSandboxModel}
{Mozilla Wiki}.
\newblock {Security/Sandbox/Process model}.
\newblock \url{https://wiki.mozilla.org/Security/Sandbox/Process_model}, 2019.

\bibitem{urlFirefoxSandbox}
{Mozilla Wiki}.
\newblock {Security/Sandbox}.
\newblock \url{https://wiki.mozilla.org/Security/Sandbox}, 2024.

\bibitem{narayan2020retrofitting}
Shravan Narayan, Craig Disselkoen, Tal Garfinkel, Nathan Froyd, Eric Rahm,
  Sorin Lerner, Hovav Shacham, and Deian Stefan.
\newblock {Retrofitting Fine Grain Isolation in the Firefox Renderer}.
\newblock In {\em USENIX Security Symposium}, 2020.

\bibitem{park2020fuzzing}
Soyeon Park, Wen Xu, Insu Yun, Daehee Jang, and Taesoo Kim.
\newblock {Fuzzing Javascript Engines with Aspect-preserving Mutation}.
\newblock In {\em IEEE Symposium on Security and Privacy (S\&P)}, 2020.

\bibitem{peng2023gleefuzz}
Hui Peng, Zhihao Yao, Ardalan~Amiri Sani, Dave~Jing Tian, and Mathias Payer.
\newblock {GLeeFuzz: Fuzzing WebGL through Error Message Guided Mutation}.
\newblock {\em USENIX Security Symposium}, 2023.

\bibitem{reis2009isolating}
Charles Reis and Steven~D Gribble.
\newblock {Isolating Web Programs in Modern Browser Architectures}.
\newblock In {\em ACM European Conference on Computer Systems (EuroSys)}, 2009.

\bibitem{rizzi2016techniques}
Eric~F Rizzi, Sebastian Elbaum, and Matthew~B Dwyer.
\newblock {On the Techniques we Create, the Tools we Build, and their
  Misalignments: A Study of KLEE}.
\newblock In {\em International Conference on Software Engineering (ICSE)},
  2016.

\bibitem{salls2021token}
Christopher Salls, Chani Jindal, Jake Corina, Christopher Kruegel, and Giovanni
  Vigna.
\newblock {Token-Level Fuzzing}.
\newblock In {\em USENIX Security Symposium}, 2021.

\bibitem{schloegel2024sok}
Moritz Schloegel, Nils Bars, Nico Schiller, Lukas Bernhard, Tobias Scharnowski,
  Addison Crump, Arash Ale-Ebrahim, Nicolai Bissantz, Marius Muench, and
  Thorsten Holz.
\newblock {SoK: Prudent Evaluation Practices for Fuzzing}.
\newblock In {\em IEEE Symposium on Security and Privacy (S\&P)}, 2024.

\bibitem{schumilo2021nyx}
Sergej Schumilo, Cornelius Aschermann, Ali Abbasi, Simon W{\"o}rner, and
  Thorsten Holz.
\newblock {Nyx: Greybox Hypervisor Fuzzing using Fast Snapshots and Affine
  Types}.
\newblock In {\em USENIX Security Symposium}, 2021.

\bibitem{segura2016survey}
Sergio Segura, Gordon Fraser, Ana~B Sanchez, and Antonio Ruiz-Cort{\'e}s.
\newblock {A Survey on Metamorphic Testing}.
\newblock {\em IEEE Transactions on Software Engineering}, 2016.

\bibitem{shou2021corbfuzz}
Chaofan Shou, Ismet~Burak Kadron, Qi~Su, and Tevfik Bultan.
\newblock {CorbFuzz: Checking Browser Security Policies with Fuzzing}.
\newblock In {\em ACM/IEEE International Conference on Automated Software
  Engineering (ASE)}, 2021.

\bibitem{srivastava2021gramatron}
Prashast Srivastava and Mathias Payer.
\newblock {Gramatron: Effective Grammar-aware Fuzzing}.
\newblock In {\em International Symposium on Software Testing and Analysis
  (ISSTA)}, 2021.

\bibitem{graphicsfuzz}
{The GraphicsFuzz Authors}.
\newblock {GraphicsFuzz Testing Framework}.
\newblock \url{https://github.com/google/graphicsfuzz}, 2019.

\bibitem{thomas2019proactive}
Gavin Thomas.
\newblock {A Proactive Approach to more Secure Code}.
\newblock
  \url{https://msrc.microsoft.com/blog/2019/07/a-proactive-approach-to-more-secure-code/},
  2019.

\bibitem{veggalam2016ifuzzer}
Spandan Veggalam, Sanjay Rawat, Istvan Haller, and Herbert Bos.
\newblock {Ifuzzer: An Evolutionary Interpreter Fuzzer using Genetic
  Programming}.
\newblock In {\em European Symposium on Research in Computer Security
  (ESORICS)}, 2016.

\bibitem{urlGPUwebW3C}
{W3C}.
\newblock {WebGPU}.
\newblock \url{https://www.w3.org/TR/webgpu/}, 2024.

\bibitem{wang2017skyfire}
Junjie Wang, Bihuan Chen, Lei Wei, and Yang Liu.
\newblock {Skyfire: Data-driven Seed Generation for Fuzzing}.
\newblock In {\em IEEE Symposium on Security and Privacy (S\&P)}, 2017.

\bibitem{wang2019superion}
Junjie Wang, Bihuan Chen, Lei Wei, and Yang Liu.
\newblock {Superion: Grammar-aware Greybox Fuzzing}.
\newblock In {\em International Conference on Software Engineering (ICSE)},
  2019.

\bibitem{wang2010taintscope}
Tielei Wang, Tao Wei, Guofei Gu, and Wei Zou.
\newblock {TaintScope: A Checksum-aware Directed Fuzzing Tool for Automatic
  Software Vulnerability Detection}.
\newblock In {\em IEEE Symposium on Security and Privacy (S\&P)}, 2010.

\bibitem{xu2020freedom}
Wen Xu, Soyeon Park, and Taesoo Kim.
\newblock {FREEDOM: Engineering a State-of-the-Art DOM Fuzzer}.
\newblock In {\em ACM Conference on Computer and Communications Security
  (CCS)}, 2020.

\bibitem{yang2011finding}
Xuejun Yang, Yang Chen, Eric Eide, and John Regehr.
\newblock {Finding and Understanding Bugs in C compilers}.
\newblock In {\em ACM SIGPLAN Conference on Programming Language Design and
  Implementation (PLDI)}, 2011.

\bibitem{zhou2023bettersemantics}
Chijin Zhou, Quan Zhang, Lihua Guo, Mingzhe Wang, Yu~Jiang, Qing Liao, Zhiyong
  Wu, Shanshan Li, and Bin Gu.
\newblock {Towards Better Semantics Exploration for Browser Fuzzing}.
\newblock In {\em ACM SIGPLAN Conference on Object-Oriented Programming
  Systems, Languages, and Applications (OOPSLA)}, 2023.

\bibitem{zhou2022minerva}
Chijin Zhou, Quan Zhang, Mingzhe Wang, Lihua Guo, Jie Liang, Zhe Liu, Mathias
  Payer, and Yu~Jiang.
\newblock {Minerva: Browser API Fuzzing with Dynamic mod-ref Analysis}.
\newblock In {\em ACM Joint European Software Engineering Conference and
  Symposium on the Foundations of Software Engineering (ESEC/FSE)}, 2022.

\end{thebibliography}
\end{document}